\documentclass{aa}  

\usepackage{graphicx}
\usepackage{txfonts}

\usepackage{amssymb}
\usepackage{amsbsy}
\usepackage{amsmath}
\usepackage{graphicx}
\usepackage{booktabs}
\usepackage{soul}
\usepackage{tabularx}
\usepackage{longtable}
\usepackage{comment}
\usepackage{siunitx}
\usepackage{verbatim}
\usepackage{natbib}
\usepackage[unicode]{hyperref}

\newcommand{\Ms}{\ensuremath{\rm M_{\odot}}}
\newcommand{\Zs}{\ensuremath{\rm Z_{\odot}}}
\newcommand{\met}{\ensuremath{\big[\rm Fe/H]}}
\newcommand{\mnickel}{\ensuremath{m_{\rm ^{56}Ni}}}
\newcommand{\nickel}{\ensuremath{ ^{56} \rm Ni}}
\newcommand{\Eejecta}{\ensuremath{E_{\rm ejecta}}}
\newcommand{\Mzam}{\ensuremath{ M_{\rm ZAMS}}}

\setstcolor{red}

\graphicspath{{./}{figures/Calibr}}
\graphicspath{{./}{figures/Plotvsini}}
\graphicspath{{./}{figures/PlotvsMCO}}

\begin{document}

    \title{The initial mass-remnant mass relation for core collapse supernovae}

   \author{Cristiano Ugolini \inst{1, 2}
          \and
          Marco Limongi\inst{2, 3, 4}
          \and
          Raffaella Schneider \inst{5, 2, 6}
          \and
          Alessandro Chieffi \inst{4, 7, 8}
          \and
          Ugo Niccolò Di Carlo \inst{1, 9}
          \and
          Mario Spera \inst{1, 2, 9}
          }

   \institute{SISSA, via Bonomea 265, 34136 Trieste, Italy\\
            \and
            Istituto Nazionale di Astrofisica - Osservatorio Astronomico di Roma, Via Frascati 33, I-00040, Monteporzio Catone, Italy \\
            \and
            Kavli Institute for the Physics and Mathematics of the Universe (WPI), The University of Tokyo Institutes for Advanced Study, The university of Tokyo, Kashiwa, Chiba 277-8583, Japan \\
            \and
            INFN. Sezione di Perugia, via A. Pascoli s/n, I-06125 Perugia, Italy\\
            \and
            Dipartimento di Fisica, Sapienza, Universita` di Roma, Piazzale Aldo Moro 5, 00185, Roma, Italy\\
            \and
            INFN, Sezione di Roma I, Piazzale Aldo Moro 2, 00185, Roma, Italy\\
            \and
            Istituto Nazionale di Astrofisica - Istituto di Astrofisica e Planetologia Spaziali, Via Fosso del Cavaliere 100, I-00133, Roma, Italy\\
            \and
            Monash Centre for Astrophysics (MoCA), School of Mathematical Sciences, Monash University, Victoria 3800, Australia\\
            \and
            National Institute for Nuclear Physics - INFN, Sezione di Trieste, I-34127 Trieste, Italy\\
             }

   \date{Received xxx; accepted xxx}

 
  \abstract
   {The first direct detection of gravitational waves in 2015 marked the beginning of a new era for the study of compact objects. Upcoming detectors, such as the Einstein Telescope, are expected to add thousands of binary coalescences to the list. However, from a theoretical perspective, our understanding of compact objects is hindered by many uncertainties, and a comprehensive study of the nature of stellar remnants from core-collapse supernovae is still lacking.}
   {In this work, we investigate the properties of stellar remnants using a homogeneous grid of rotating and non-rotating massive stars at various metallicities from \citet{limongi_presupernova_2018}.}
   {We simulate the supernova explosion of the evolved progenitors using the HYdrodynamic Ppm Explosion with Radiation diffusION (HYPERION) code \citep{limongi_hydrodynamical_2020}, assuming a thermal bomb model calibrated to match the main properties of SN1987A.}
   {We find that the heaviest black hole that can form depends on the initial stellar rotation, metallicity, and the assumed criterion for the onset of pulsational pair-instability supernovae. Non-rotating progenitors at $\met=-3$ can form black holes up to $\sim 87,\Ms$, falling within the theorized pair-instability mass gap. Conversely, enhanced wind mass loss prevents the formation of BHs more massive than $\sim 41.6,\Ms$ from rotating progenitors. We use our results to study the black hole mass distribution from a population of $10^6$ isolated massive stars following a Kroupa initial mass function. Finally, we provide fitting formulas to compute the mass of compact remnants as a function of stellar progenitor properties. Our up-to-date prescriptions can be easily implemented in rapid population synthesis codes.}
   {}


\titlerunning{Remnant masses for core collapse supernovae}
\authorrunning{C.Ugolini, M. Limongi, R. Schneider, A. Chieffi, U. N. Di Carlo, M. Spera}
   
   \maketitle
%
\section{Introduction} \label{sec:intro}

On September 14, 2015 the LIGO interferometers captured a gravitational-wave (GW) signal from two merging black holes (BHs) with masses $\sim 36 \,\Ms$ and $ \sim 29 \,\Ms$, respectively \citep[]{abbott_gw150914_2016, abbott_observation_2016, LIGO_AstroImpl_2016}.
The event, named GW150914, was the first direct detection of GWs, and the proof that BH binaries (BHB) exist and can merge within a Hubble time. Furthermore, prior to GW150914, there were no unambiguous evidences for stellar BHs more massive than $ \sim 25 \,\Ms$, which is the upper limit for the masses of BHs in known X-ray binaries(\citealt{orosz2002inventory, Oezel2010, MillerJones2021}).
Thus, GW150914 also provides a robust evidence for the existence of heavy stellar BHs.
To date, more than 90 BH-BH mergers have been reported by the LIGO-Virgo-KAGRA (LVK) collaboration, \citep{O1_O2_LIGO_2019, O3a_2021, O3b_LIGO_2021} and in about 70 of the events, at least one of the two BHs has mass $\gtrsim 25 \Ms$. Moreover, the number of detections is expected to significantly increase in the next years because of the recent O4 LVK observing run\footnote{See  \href{https://observing.docs.ligo.org/plan/}{LIGO, Virgo AND KAGRA observing run plans}.} and of the upcoming next-generation detectors (e.g., Einstein Telescope and Cosmic Explorer). \par
From the theoretical point of view, the formation and the evolutionary pathways of BHs and BHB are very uncertain \citep[e.g. ][]{Mapelli2021, Spera2022}.
Two main astrophysical formation channels have been proposed: the isolated binary evolution scenario and the dynamical formation pathway. In the former, the stellar progenitors evolve in complete isolation as a gravitationally bound pair and they eventually turn into compact objects and merge within a Hubble time \citep[see][]{Heuvel1973, Heuvel2017, Heggie1975, Tutukov1993, Brown1995, Bethe1998, Sandquist1998, Dewi2006, Justham2011, Dominik2012, Dominik2015, Schneider2015, Belczynski2016, Belczynski2018, Marassi2019, Belczynski2020, Pavlovskii2017, Gomez2018, Wysocki2018, Mapelli2018, Neijssel2019, Graziani2020, Mandel2020a, Mandel2022, vanSon2020, Broekgaarden2021, Gallegos-Garcia2021, Marchant2021, Zevin2021}. In the latter, a compact-object binary forms and shrinks after a series of gravitational few-body interactions in dense stellar environments \citep[see][]{Sigurdsson1993, Sigurdsson1995, Moody2009, Banerjee2010, Banerjee2017, Banerjee2018, Banerjee2021, Downing2011, Mapelli2013, Mapelli2016, Ziosi2014, Rodriguez2015, Rodriguez2016, Rodriguez2019, DiCarlo2019, DiCarlo2020, DiCarlo2021a, DiCarlo2021b, Rastello2019, Rastello2020, Rastello2021, Kremer2020a, Wang2020, Wang2022, Ye2022}.\par
However, both the isolated and the dynamical scenarios rely on many uncertain astrophysical processes, including the  core-collapse supernova (ccSN) mechanism. Our knowledge of the SN engine is still limited, thus investigating the explosion mechanism, the stellar explodability, and the mass spectrum of stellar remnants is challenging.
All stars with Zero-Age Main Sequence (ZAMS) mass $\left(\Mzam\right) > 9 \,\Ms$ \citep{Roberti2024} are expected to end their life with a ccSN and form a compact object that can be either a BH or a neutron star (NS), depending on the mass of the stellar progenitor and on the ongoing binary processes (see e.g. \citealt{Doherty2017, Poelarends2017, Zapartas2017}).

The current reference model for ccSNe is based on the neutrino driven explosion \citep[]{Bethe1985, Burrows1995, Janka_1996, Janka2016, Janka2017, Kotake2006, Herant1994, Burrows1995, Janka_1996, Foglizzo_2006, Kotake2006, Burrows2019, Aguilera-Dena2022}. 

State-of-the-art 3D simulations of the explosion, which include the most sophisticated treatment of neutrino transport, cannot obtain naturally and in a routinely way the explosion of the star. In addition, such simulations cannot accurately constrain the amount of matter that can fall back onto the proto-compact object at later times and, consequently, the final mass of the compact remnant \citep[see][]{Blondin_2003,Janka_1996, Burrows1995, Suwa2010, Suwa2013, Muller2012, Muller2015, Summa2016, Bruenn2016, Lentz_2012, Lentz2015, Burrows2019, Burrows2020, Mezzacappa2020, Bollig_2021, Wang2020, Wang2022, Burrows_2023, burrows2024theoryneutronstarblack, Vartanyan2023, Mezzacappa_2023, Powell_2023, Kinugawa2021, kinugawa2023, nakamura2024, Shibagaki_2024}.

A simplified approach to study the explosion of massive stars is to artificially stimulate the SN explosion by injecting some amount of kinetic energy (i.e., a kinetic bomb, see e.g. \citealt{Limongi2003, Chieffi2004}) or thermal energy (i.e., a thermal bomb, see e.g. \citealt{Thielemann1996}) at an arbitrary mass coordinate location of the progenitor model.
Alternatively, the inner edge of the exploding mantle is moved outward like a piston following the trajectory of a projectile launched with a specific velocity in a given potential (see \citealt{Woosley1995}). 
Independently of the adopted techniques, the evolution of the shock is followed by means of 1D hydrodynamical simulations.
The main advantage of this approach is that the computing cost is relatively small. Thus, it is possible to follow the evolution of the shock over longer timescales and to estimate the amount of fallback and the final remnant mass. 
The 1D approach has been extensively used to study the SN explosion yields, especially in the context of explosive nucleosynthesis \citep{Shigeyama1988, Hashimoto1989, Thielemann1990, Thielemann1996, Nakamura2001, Nomoto2006, Umeda2008, Moriya2010, Bersten2011, chieffi_pre-supernova_2013, limongi_presupernova_2018}.\par

Through 1D simulations, several authors have attempted to link the explodability of stellar progenitors to the structural properties of stars at the presupernova (pre-SN) stage \citep{oconnor_black_2011, fryer_compact_2012, Horiuchi2014, ertl_two-parameter_2016, sukhbold_core-collapse_2016, Mandel2020b, Patton2020, Patton2022}. Such prescriptions for the explodability have been adopted in many population synthesis codes (e.g.,  \citealt{Eldridge2008, Spera2015, Agrawal2020, Zapartas2021, Fragos2023}), to predict the final fate of massive stars and the formation and evolution of compact-object binaries in different astrophysical environments. Such analytical relations for the explodability are usually applied to stellar progenitors that are different from those employed to compute the prescriptions, and therefore must be adopted with caution. In addition, we are still far from making any conclusive statement on the final fate of massive stars. Specifically, the role played by stellar rotation, metallicity, and different stellar evolutionary models of the progenitors is still mostly unexplored.\par
In this work, we try to address some of these problems by studying the explosion of progenitor stars and the formation of compact remnants through the latest version of our hydrodynamic code HYPERION (HYdrodynamic Ppm Explosion with Radiation diffusION, see \citealt{limongi_hydrodynamical_2020}), which includes a treatment of the radiation transport in the flux-limited diffusion approximation. We simulate the explosion of an up-to-date homogeneous set of both rotating and non-rotating progenitor stars provided by \citet[][hereinafter LC18]{limongi_presupernova_2018}.
For the SN explosion, we adopt the thermal-bomb approach and we calibrate the parameters that describe the energy-injection process by comparison with SN1987A.
We apply our methodology to investigate the distribution of remnant masses as a function of $\Mzam$,  focusing on the role played by different initial metallicities and different values of initial stellar rotation.\par
In Section \ref{sec: Methods}, we present our methodology. In Section \ref{sec: Results} we present the results of our simulations.
In Section \ref{sec: Discussion} we investigate the implications of our results, and Section \ref{sec: summary and conclusions} contains the conclusions.

\section{Methods}\label{sec: Methods}

We use the HYPERION code to simulate the explosion of a set of pre-SN models of massive stars presented in LC18. The details of the code are discussed in Sec. \ref{sec: Algorithm}.\par
To induce the explosion, we remove the innermost $0.8\,\Ms$ of the pre-SN model, and artificially deposit at this mass coordinate a given amount of thermal energy ($E_{\rm inj}$).
We assume that $E_{\rm inj}$ is diluted both in space, over a mass interval $dm_{\rm inj}$,  and in time, over a time interval $dt_{\rm inj}$.
Therefore, our numerical approach relies on the three free parameters  $E_{\rm inj}$, $dm_{\rm inj}$, and $dt_{\rm inj}$, which need to be calibrated. The details of the calibration are presented in Sec. \ref{sec: Calibration of the explosion}. \par
Once the energy is deposited into the pre-SN model, a shock wave forms. The shock wave starts to propagate outwards and eventually drives the explosion of the star. While the shock advances, it induces explosive nucleosynthesis and accelerates the shocked matter. After the initial energy injection, the innermost regions begin to fall back onto the proto-compact object, which progressively increases its mass. After the end of the fallback phase, we define the \textit{mass cut} as the mass coordinate that divides the final remnant from the ejecta. Depending on the position of the mass cut, we can recognize two possible outcomes of our simulations: successful explosion, if a substantial fraction of the envelope is ejected during the explosion (the simulated physical time is of 1 yr, $\sim 3\times 10^7 s$, which is sufficient to follow also the homologous expansion of the ejected matter in the circumstellar medium and the formation of the light-curve), or failed explosion if essentially the whole star collapses to a black hole. The former case is presented in Sec. \ref{sec: SN run}, while the latter is discussed in Sec. \ref{sec: failed-SN run}.

\subsection{The HYPERION code} \label{sec: Algorithm}
In this section, we briefly describe HYPERION, which is the 1D hydrodynamical code we adopt in this work and which was presented in detail in \cite{limongi_hydrodynamical_2020}.
The code solves the full system of hydrodynamic equations (written in a conservative form) supplemented by the equation of the radiation transport (in the flux-limited diffusion approximation) and the equations describing the temporal variation of the chemical composition due to the nuclear reactions. Such system of equations can be divided into two parts. The first part includes the hydrodynamical equations and the radiation transport:

\begin{align}
	\dfrac{\partial \rho}{\partial t}&=-4\pi\rho^2 \dfrac{\partial r^2v}{\partial m}\\[11 pt]
	\dfrac{\partial v}{\partial t}   &=-4\pi r^2\dfrac{\partial P}{\partial m}-\dfrac{Gm}{r^2}\\[12 pt]
	\dfrac{\partial E}{\partial t}   &=-\dfrac{\partial }{\partial m}\bigg(4\pi r^2vP+L\bigg)+\epsilon
    \label{sist: Hydro}
\end{align}
where $\rho$ is the density, $r$ is the radial coordinate, $v$ is the velocity, $m$ is the mass coordinate, $P$ is the pressure, $E$ is the total energy per unit of mass, $L$ is the radiative luminosity, and $\epsilon$ is any source and/or sink of energy (e.g. nuclear reactions, neutrino losses).\par
The second part describes the variation of chemical composition caused by nuclear reactions:
\begin{equation}
    \begin{split}
        \dfrac{\partial Y_i}{\partial t} \bigg|_{i=1, ...N}= &\sum_j c_i(j)\Lambda_jY_j\\
        &+\sum_{j,k}c_i(j,k)\rho N_A \langle \sigma v\rangle_{j,k}Y_jY_k \\
        &+\sum_{j, k, l}c_i(j, k, l)\rho^2 N_A^2 \langle \sigma v\rangle_{j,k, l}Y_jY_kY_l \end{split}
    \label{eq: Chem Evo}
\end{equation}
where $N$ is the number of nuclear species included in the calculations, $Y_i$ is the abundance by number of the $i$-th nuclear species, $\Lambda$ refers to the weak interactions or the photodissociation rate, depending on which process is in place for the $j$ nuclear species, and $\langle \sigma\bar{v}\rangle$ represents the two or three-body nuclear cross section (see \citealt{limongi_presupernova_2018} and \citealt{limongi_hydrodynamical_2020} for a detailed discussion).\par
The terms on the right side of Eq. \ref{eq: Chem Evo} describe the contribution to the abundance  due to various physical processes, such as: (1) the $\beta$-decay, electron-captures, and photodisintegrations; (2) two-body reactions; (3) three-body reactions. The coefficients in Eq. \ref{eq: Chem Evo} are defined as:
\begin{align}
    c_i(j)=&\pm N_i\\
    c_{i}(j,k)=& \pm\dfrac{N_i}{(N_j)!(N_k)!}\\
    c_{i}(j,k,l)=& \pm\dfrac{N_i}{(N_j)!(N_k)!(N_l)!}
\end{align}
where $N_i$ refers to the number of $i$ particles involved in the reaction, and the factorial terms prevent from double counting the reactions with identical particles. The sign (+)/(-) depends on whether the $i$ particles are created or destroyed.
Finally, the radiative luminosity in the flux-limited diffusion approximation is defined as:
\begin{equation}
L=-(4\pi r^2)^2 \dfrac{\lambda ac}{3\kappa}\dfrac{\partial T^4}{\partial m},
\label{eq: Luminosity}
\end{equation}
where $a$ is the radiation constant, $c$ is the speed of light, $\kappa$ is the Rosseland mean opacity, and $\lambda$ is the flux limiter, computed according to \cite{Levermore1981} as:
\begin{equation}
\lambda=\dfrac{6+3R}{6+3R+R^2},
\end{equation}
with
\begin{equation}
	R=\dfrac{4\pi r^2}{kT^4}\bigg|\dfrac{\partial T^4}{\partial t}\bigg|.
	\label{eq: Parametro R}
\end{equation}
The Rosseland mean opacities are calculated for a scaled solar distribution of all elements, corresponding to $Z=Z_\odot=1.345\times10^{-2}$ ($\met=0$) \citep{Asplund2009}. At lower metallicities, we consider an enhancement of C, O, Mg, Si, S, Ar, Ca, and Ti with respect to Fe, accordingly to observations \citep[][]{Cayrel2004, Spite2005}. As a result of the enhancement, we find that metallicity indexes $\met=-1,\,-2$ and $-3$, correspond to metallicity values of $Z=3.236\times10^{-3}$ ($2.406\times10^{-1}\, \Zs$), $Z=3.236\times10^{-4}$ ($2.406\times10^{-2}\, \Zs$), and $Z=3.236\times10^{-5}$ ($2.406\times10^{-3}\, \Zs$), respectively.\par
We adopt opacity tables from three different sources, depending on the temperature regime. In the low-temperature regime ($2.75\leq\log T/{\rm K} \leq 4.5$), we refer to \cite{Ferguson_2005}, in the intermediate-temperature regime ($4.5\leq\log T/{\rm K} \leq 8.7$) to \cite{Iglesias1996}, and for higher temperatures we adopt the Los Alamos Opacity Library \citep{Huebner_1977}.\par
Lastly, the equation of state (EOS) is the one described in \cite{Morozova2015}, based on the analytic EOS of \cite{Paczynski1983}, which takes into account radiation, ions, and electrons in an arbitrary degree of degeneracy. The H and He ionization fractions are obtained by solving the Saha equations, as in \cite{Zaghloul2000}, while we assume that all other elements are fully ionized.\par
The energy generated through nuclear reactions is neglected, because we assume that its contribution is negligible compared to other energy sources, but we consider the energy deposition from $\gamma$-rays that comes from the radioactive decays $\rm ^{56}Ni\longrightarrow ^{56}Co + e^+ +\nu_e + \gamma$ and $\rm ^{56}Co\longrightarrow ^{56}Fe + e^+ +\nu_e + \gamma$, following the scheme described in \cite{Swartz1995} and \cite{Morozova2015}.\par
To solve the hydrodynamic equations, we use a fully Lagrangian scheme of piece-wise parabolic method (PPM) presented in \cite{Colella1984}. To implement that, we first interpolate the profiles of the variables $\rho,\,v,\,\text{and}\,P$ as a function of the mass coordinate, using the interpolation algorithm of \cite{Colella1984}. Then, 
we solve the Riemann problems at every cell interfaces. Thus, we obtain the time-averaged values of pressure and velocity at the zone edges.
Finally, we update the conserved quantities by applying the forces resulting from the time-averaged pressures and velocities at the zone edges, following the scheme presented in \cite{limongi_hydrodynamical_2020}.

\subsection{Calibration of the explosion}\label{sec: Calibration of the explosion}

During the explosion, the innermost zones of the stellar mantle are heated up by the expanding shock wave to temperatures high enough to induce explosive nucleosynthesis. One of the main products of the nucleosynthesis is $\nickel$, which is synthesized by the explosive Si burning in the innermost regions of the mantle \citep[][]{Thielemann1996, Woosley1995, Arnett1996, Limongi2000}. Therefore, the location of the mass cut is crucial to constrain the amount of ejected $\nickel$: in absence of mixing, the more external the mass cut, the lower the abundance of nickel in the ejecta.
Another important outcome of our simulations is the final kinetic energy of the ejecta, which comes from the conversion of a fraction of the thermal energy initially injected into the progenitor star. 
The amount of ejected $\nickel$ and the final kinetic energy of the ejecta depend on the explosion parameters, i.e. $E_{\rm inj}$, $dm_{\rm inj}$, and $dt_{\rm inj}$. We calibrate these parameters considering SN1987A, which is the most extensively studied ccSN to-date \citep[e.g.][]{Arnett1989, Woosley1988}. Specifically, our goal is to find the combination of the explosion parameters that match both the ejected $\nickel$ and the kinetic energy of the ejecta with the values estimated for the SN 1987A, i.e.  $m_{\nickel}\sim 0.07 \, \Ms$ and $\sim10^{51}\,\text{erg}=1 \, \text{foe}$ (see, e.g., \citealt{Arnett1989, Shigeyama1990, Utrobin1993, Utrobin2006, Blinnikov2000}). 
To reach our goal, we simulate the explosions of the stellar progenitor of our grid that best match the helium core of the progenitor of SN 1987A (i.e., Sk — 69°202 e.g. \citealt{Woosley1988, Arnett1989, Arnett1996} with $M_{\rm He}\sim 6\,\Ms$, e.g. \citealt{Woosley1988}).
We made this choice because $\Eejecta$ and $m^{\nickel}$ do not depend on the structure of the hydrogen envelope of the star.\par
Therefore, we adopt a $15 \, \Ms$-progenitor model with initial solar composition ($\met=0$, i.e. the model 15a000 in the set of LC18) and, in each simulation, we adopt a different set of explosion parameters.
\begin{figure}[htb]
    \centering
    \includegraphics[width=\columnwidth]{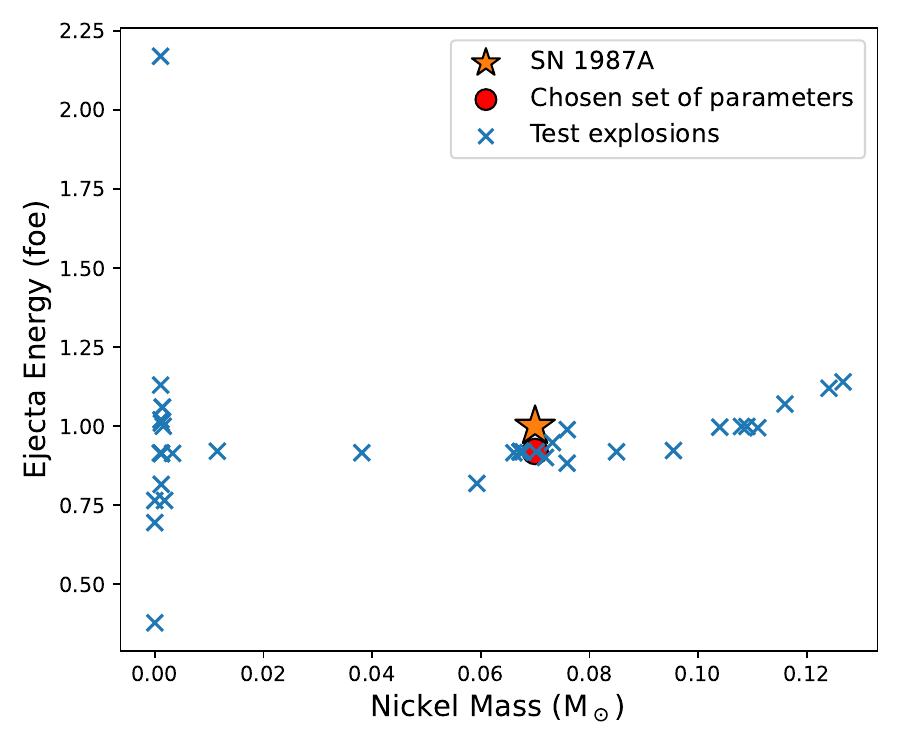}
    \caption{Final kinetic energy of the ejecta as a function of the ejected mass of  $\nickel$ obtained running various explosion tests (blue crosses) with different values of the initial explosion parameters, $E_{\rm inj}$, $dm_{\rm inj}$, and $dt_{\rm inj}$. The red dot refers to the simulation of the explosion that best matches the properties of SN1987A (orange star).}
    \label{fig: Cal parameters}
\end{figure}
Figure \ref{fig: Cal parameters} shows the final kinetic energy of the ejecta and the corresponding amount of ejected $\nickel$ obtained from our simulations using different explosion parameters, with $E_{\rm inj} \in \left[1.9 ; 2.1\right]\, \rm foe$, $dm_{\rm inj} \in \left[0.05  ; 0.4\right] \, \Ms$, and $dt_{\rm inj} \in \left[ 10^{-9}  ;  1\right] \, \rm s$.

The combination of explosion parameters that best matches the values of SN1987A is: $dt_{\rm inj}=0.01\, s$, $dm_{\rm inj} = 0.1 \, \Ms$, and $E_{\rm inj} = 2.0 \, \rm foe$. We adopt this combination of explosion parameters in all the simulations presented in this work. Using the calibrated values of $E_{\rm inj}$, $dm_{\rm inj}$, and $dt_{\rm inj}$, we compute the explosion of a subset of the pre-SN models presented in LC18.

Finally, it is important to mention a significant caveat that may influence these results. \citet{Podsiadlowski_1992} and \citet{Menon_2017} suggest that Sk — 69°202, the progenitor of SN1987A, might have originated from a stellar merger. This scenario introduces uncertainties in the final He core mass of the progenitor and, consequently, in the dynamics of the explosion. However, SN1987A is widely used in the literature (e.g., \citealt{Ugliano2012, ertl_two-parameter_2016, sukhbold_core-collapse_2016, Sukhbold2018}) as a benchmark to tune the parameter describing the supernova explosion mechanism. Adopting this approach remains essential for facilitating comparisons with other studies.

\subsection{Progenitors unstable against pair production} \label{sec: PI regime}

\begin{figure*}[htb!]
    \centering
    \includegraphics[width=\textwidth]{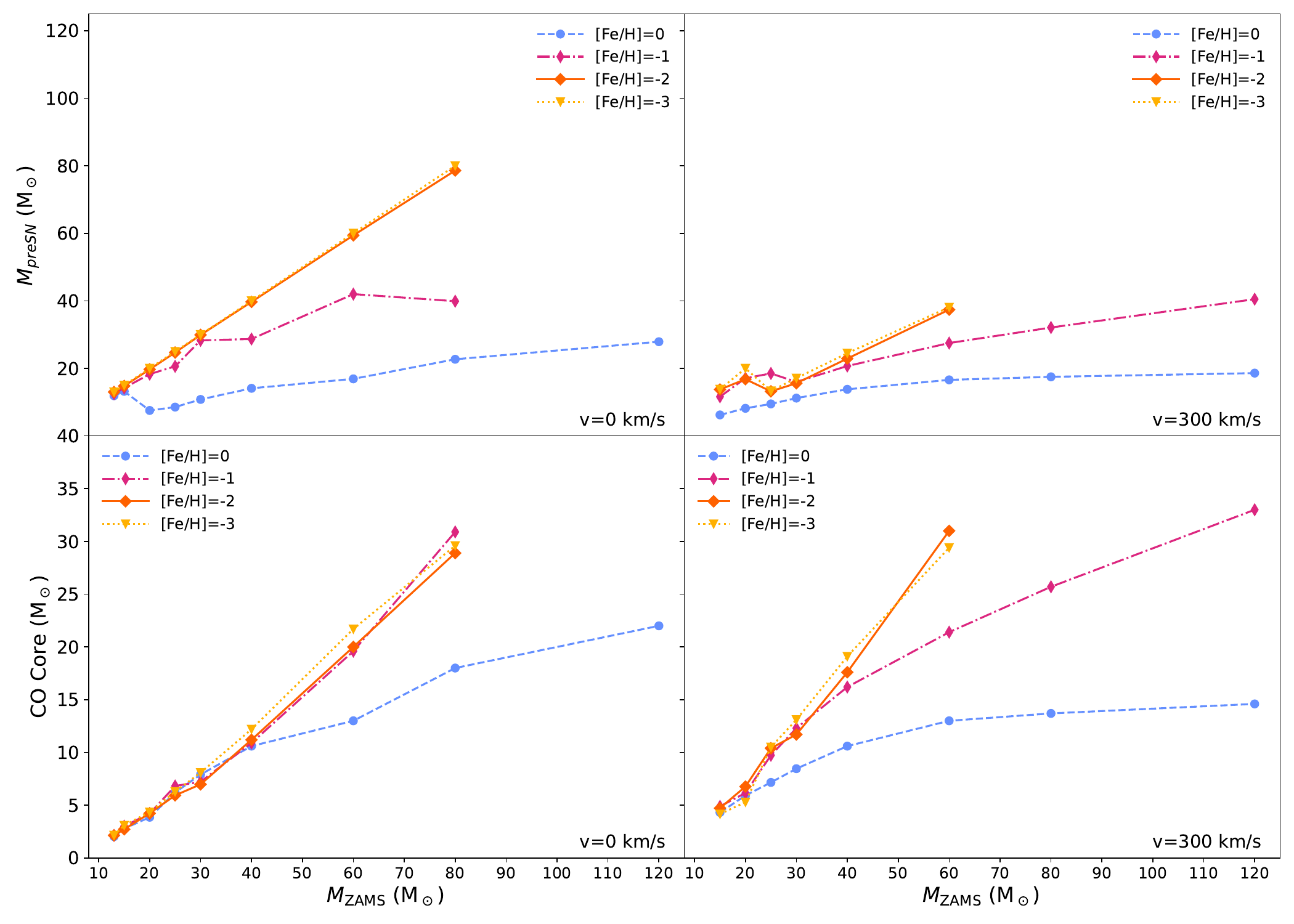}
    \caption{Mass at the pre-SN stage (top-row panels) and CO core mass (bottom-row panels) as a function of $\Mzam$ for the non-rotating progenitors (left panels) and progenitors rotating with an initial velocity of $300\,\rm km/s$ (right panels). Different colors represent different initial metallicities: $\met=0$ (dashed light blue line), $\met=-1$ (dash-dotted violet line), $\met=-2$ (orange solid line), and $\met=-3$ (dotted yellow line).}
    \label{fig: plot co and mpresn vs mzams}
\end{figure*}

Figure \ref{fig: plot co and mpresn vs mzams} shows the mass at the pre-SN stage $M_{\rm preSN}$ and the CO core mass $M_{\rm CO}$ of the stars of our simulations grid as a function of $\Mzam$. The properties of the models for which we perform the explosions are also summarized in Table \ref{Tab: Non rotating simulations} and Table \ref{Tab: Rotating simulations}.

In our grid of non-rotating stellar models, we find that at subsolar metallicities the non-rotating stellar progenitors with $\Mzam=80\,\Ms$ end their life as ccSNe with CO core masses of $\sim 30 \,\Ms$, while the $120\,\Ms$-progenitors become unstable (i.e., the adiabatic index $\Gamma_{1}$ approaches the critical value of $4/3$ in a substantial fraction of the core) and start entering the pair-instability phase (\citealt{Fowler1964, Barkat1967, Rakavy1967}). 
Thus, the progenitors unstable against pair production start to evolve on the hydrodynamical timescale and the FRANEC code cannot follow the evolution of the stellar structure anymore. Furthermore, pulsational pair-instability supernovae 
(PPISNe) are very different from the ccSNe HYPERION is designed to deal with, thereby we cannot simulate the final stages of massive stars that develop pair instabilities.\par
Finally, the mass resolution of the pre-SN simulated grid does not allow us to determine the mass threshold for the onset of the PPISNe; from our data, the only thing we know is that $120\,\Ms$ progenitors at sub-solar metallicity are unstable against pair production, while the $80\,\Ms$ progenitors are not. In order to determine which is the minimum mass of the CO core that enters the PPISNe regime ($\rm M^{\rm PPISN}_{\rm CO}$) and the minimum one that enters the PISNe ($\rm M^{\rm PISN}_{\rm CO}$), we can refer to the values provided by \cite{Heger2002} (hereinafter HW02) and \cite{Woosley2017} (hereinafter W17), i.e. $\rm M^{\rm PPISN}_{\rm CO, HW02}=33\, \Ms$ and $\rm M^{\rm PPISN}_{\rm CO, W17}=28\,\Ms$, respectively. These two values are in a good agreement with the LC18 models ($\rm M^{\rm PPISN}_{\rm CO, LC18}>33\, \Ms$), within the theoretical uncertainties. Additionally, it should be noted that, in W17, the pulses associated with a CO core mass of $ M_{\rm CO}=28-31 \, \Ms$ are quite weak, and the first significant pulse occurs at a CO core mass of approximately $33\,\Ms$.  
Hence, we adopt $\rm M^{\rm PPISN}_{\rm CO, W17} = 33\,\Ms$ as the minimum CO core mass that enters the PPISN. Furthermore, since the liming masses obtained in both cases (i.e. HW02 and W17) are compatible with the results of LC18, in the following we will refer, and use, the results provided by W17.
Finally, following W17, stars with CO core mass $\geq 54\,\Ms$ are completely disrupted by PISNe. This value corresponds to an initial mass which is larger than the maximum mass considered in our SN grid.\par
Independently of the rotation rate and metallicity, Figure \ref{fig: plot co and mpresn vs mzams} shows that $M_{\rm preSN}$ and $M_{\rm CO}$ scale linearly with $\Mzam$.\par
Therefore, to find the values of $\Mzam$ and $M_{\rm preSN}$ of the first progenitor unstable against pair instabilities we assume:

\begin{equation}
   \Mzam^{\rm PPISN}=\big(\mathrm{ M_{CO}^{PPISN}}-q_1\big)/ k_1  \label{eq: Mzams interpolation}    
\end{equation}

\begin{equation}
   \begin{split}
    M_{\rm preSN}^{\rm PPISN}&= k_2\Mzam^{\rm PPISN}+q_2 \label{eq: Mpresn interpolation}\\
    &=\dfrac{k_2}{k_1}\big( {\rm M_{CO}^{PPISN}}-q_1\big)+q_2 
   \end{split} 
\end{equation}

where $k_1$ and $q_1$ ($k_2$ and $q_2$) are the slope and intercept of the line $ M_{\rm CO}$ vs $\Mzam$ ($ M_{\rm preSN}$ vs $\Mzam$).

Therefore, according to both W17 we find $\Mzam^{\rm PPISN}$ for the various metallicities and rotation velocities that are reported in Table \ref{Tab: Interpolation results}. Specifically, in Table \ref{Tab: Interpolation results} we report, for all the initial values of metallicity and angular rotation, the values of $\Mzam^{\rm PPISN}$ and $M_{\rm preSN}^{\rm PPISN}$ that we obtain for the last progenitors stable against pair-instabilities. When the heaviest stars of our pre-SN grid are stable against pair-production we cannot predict for which values of $\Mzam$ they will become unstable against pair-instability (i.e. which stellar progenitors will explode as PPISNe or PISNe), thus $\Mzam^{\rm PPISN}$ and $M_{\rm preSN}^{\rm PPISN}$ are labelled as Not Available (N.A.) in Table \ref{Tab: Interpolation results}.

\section{Results} \label{sec: Results}

 In this section, we present the results of the simulations performed in the pre-SN models described in Figure 2 and Tables 2 and 3 (see also LC18).

We simulate the explosions for all the models with pre-SN CO core masses $\leq 33 \, \Ms$ because these models are
considered stable against pair production, as in LC18 (see Section \ref{sec: PI regime}).
The results of the simulations are also summarized in Tables \ref{Tab: Non rotating simulations} and \ref{Tab: Rotating simulations}.

\subsection{Successful explosions} \label{sec: SN run}

\begin{figure}[htb]
    \centering
    \includegraphics[width=\columnwidth]{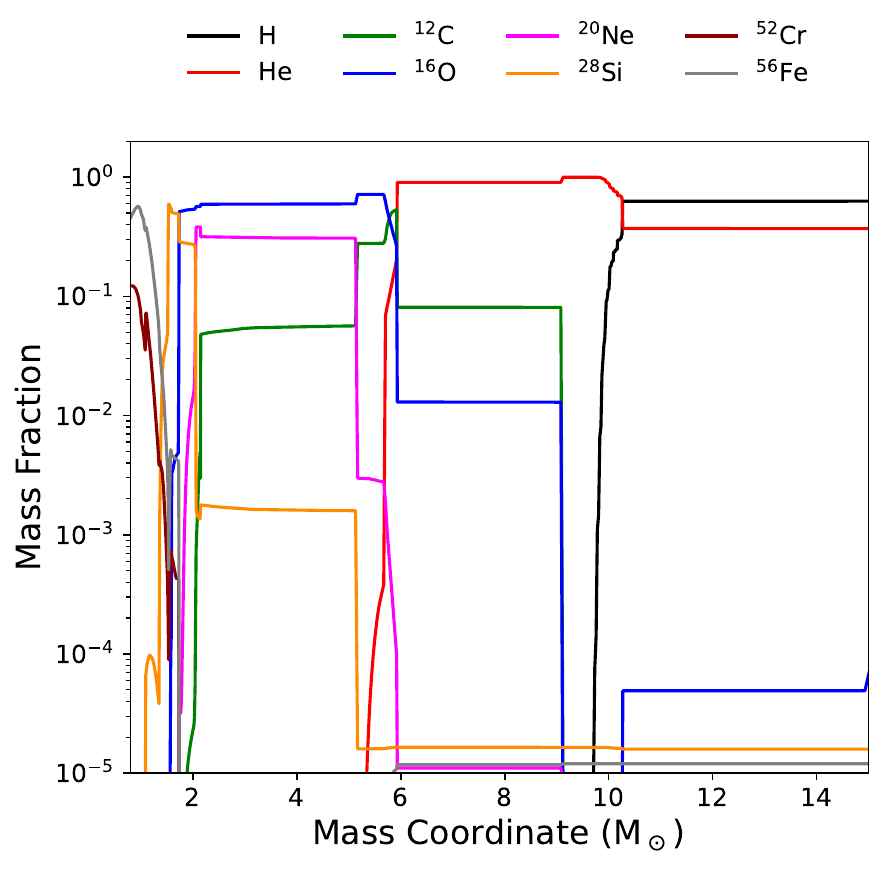}       
    \caption{
    Chemical composition as a function of the interior mass coordinate of a non-rotating model with $\Mzam=25\,\Ms$ and $\met=-2$ at the pre-SN stage. We show the chemical composition up to the mass coordinate of $15\,\Ms$. We report the abundances of H (black line), He (red line), $^{12}C$ (green line), $^{16}O$ (blue line), $^{20}Ne$ (magenta line), $^{28}Si$ (orange line), $^{52}Cr$ (dark red line) and $^{56}Fe$ (gray line).}
    \label{fig: Chemical Composition 25Ms progenitor}
\end{figure}

To illustrate the properties of a typical successful explosion in our simulations, we consider a typical successful explosion looks like in our simulations, we consider a non-rotating $25 \, \Ms$ star with metallicity $\met=-2$. Figure \ref{fig: Chemical Composition 25Ms progenitor} shows the chemical structure of the model which is characterized by an extended H-rich envelope, with a He core mass of $9.87\, \Ms$ and a CO core mass of $5.95 \, \Ms$. The ONe shell extends from mass coordinate $1.73 \, \Ms$ to $5.17 \, \Ms$, while the Si shell, which is enriched by the products of the O-shell burning, is in the range $1.43 \, \Ms$ --- $ 1.73 \, \Ms$. The mass of the iron core is therefore $1.43 \Ms$.\par
\begin{figure}[htb]
    \centering
    \includegraphics[width=\columnwidth]{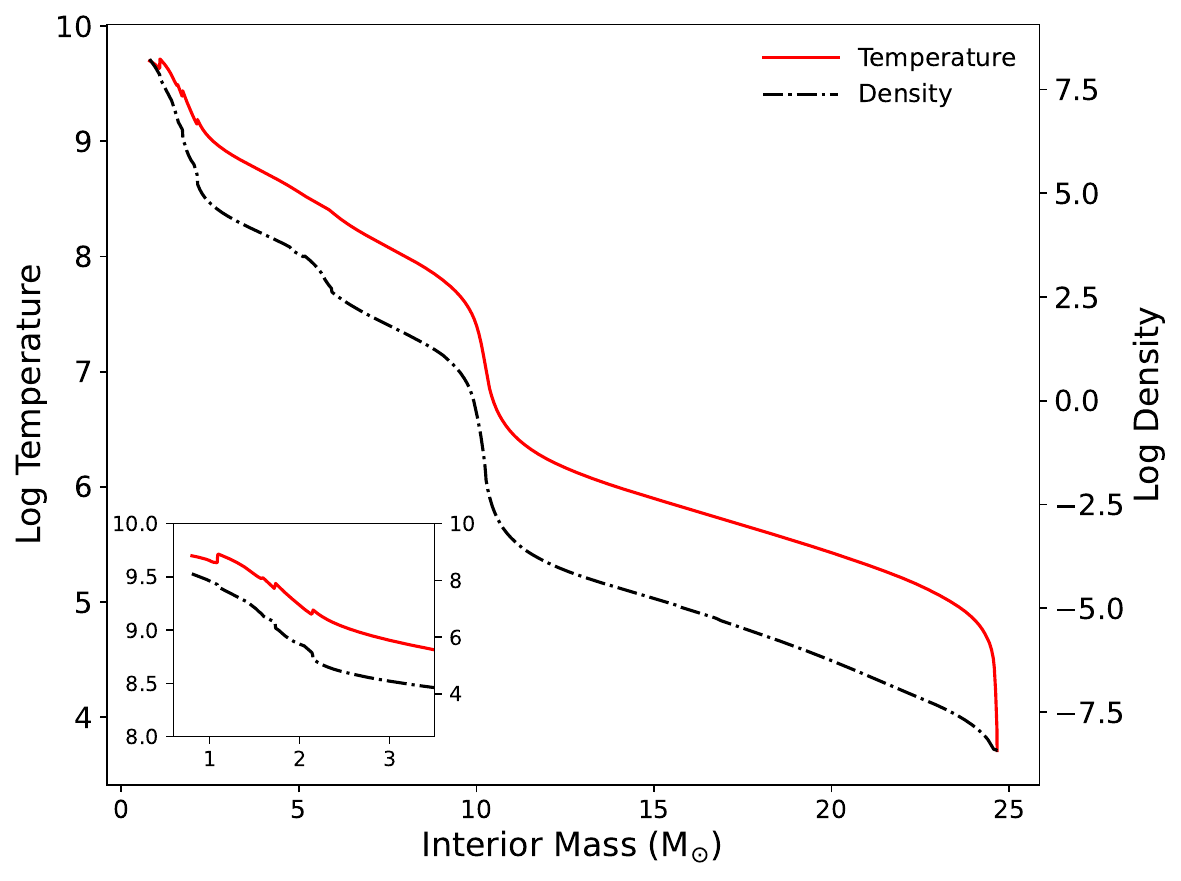}
    \caption{Temperature profile (solid red line) and mass density profile (dash-dotted black line) of the stellar progenitor at the pre-SN stage as a function of the mass coordinate. A zoom-in on the innermost region of the core, up to $5\,\Ms$, is highlighted in the lower right box.}
    \label{fig: Physical Composition 25Ms progenitor}
\end{figure}

Figure \ref{fig: Physical Composition 25Ms progenitor} shows the temperature and density profiles of the star at the pre-SN stage. 
During the pre-SN evolution, the progenitor loses a negligible amount of mass because of its initial low metallicity and has a low effective temperature of $\sim 4.7\times 10^3 \, \rm K$.
Thus, it ends its life as a Red Super Giant (RSG, \citealt{limongi_presupernova_2018}). 


\begin{figure}[htb]
    \centering
    \includegraphics[width=\columnwidth]{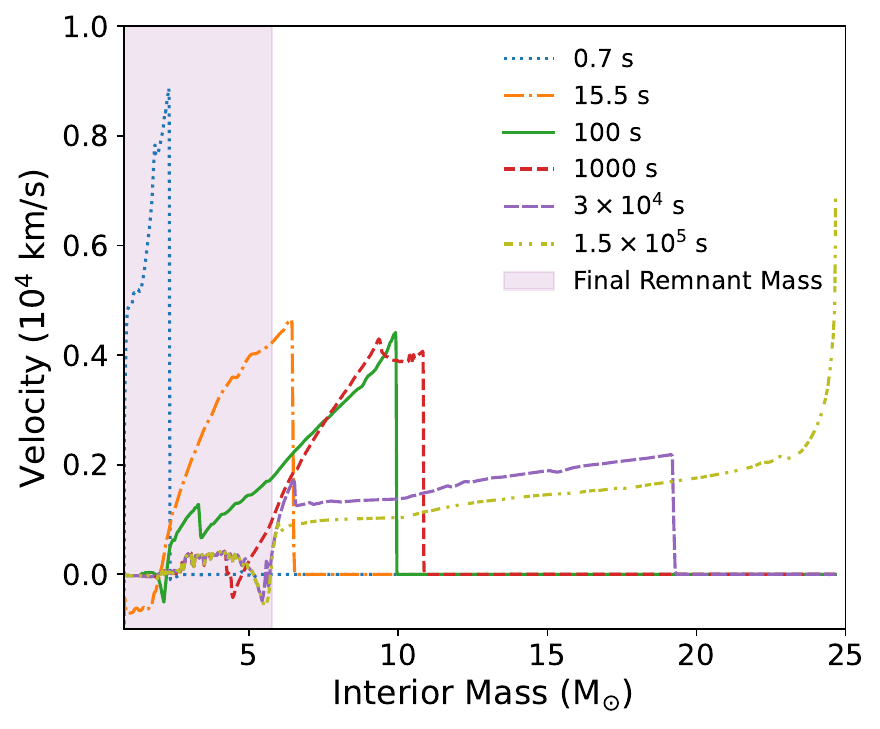}
    \caption{Velocity of the shock wave as a function of the mass coordinate at different times after the onset of the explosion: the end of the explosive nucleosynthesis (dotted blue line), the shock wave enters the CO core (dash-dotted orange line), the shock wave reaches the He/H interface (solid green line), 1000 s after the onset of the explosion (dashed red line), the formation of the compact remnant (long dashed purple line), and the shock breakout (short dot-dashed light green line).     
    The shaded purple area represents the extension of the final compact remnant.}
 
    \label{fig: SN vel}
\end{figure}

\begin{figure*}[htb]
    \centering
    \includegraphics[width=\columnwidth]{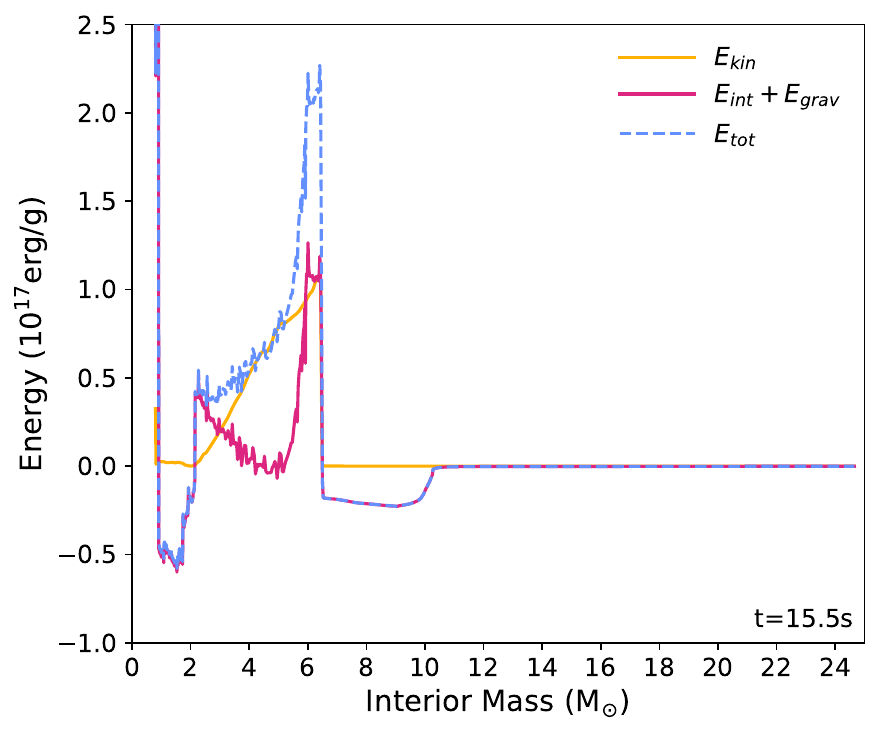}\quad
    \includegraphics[width=\columnwidth]{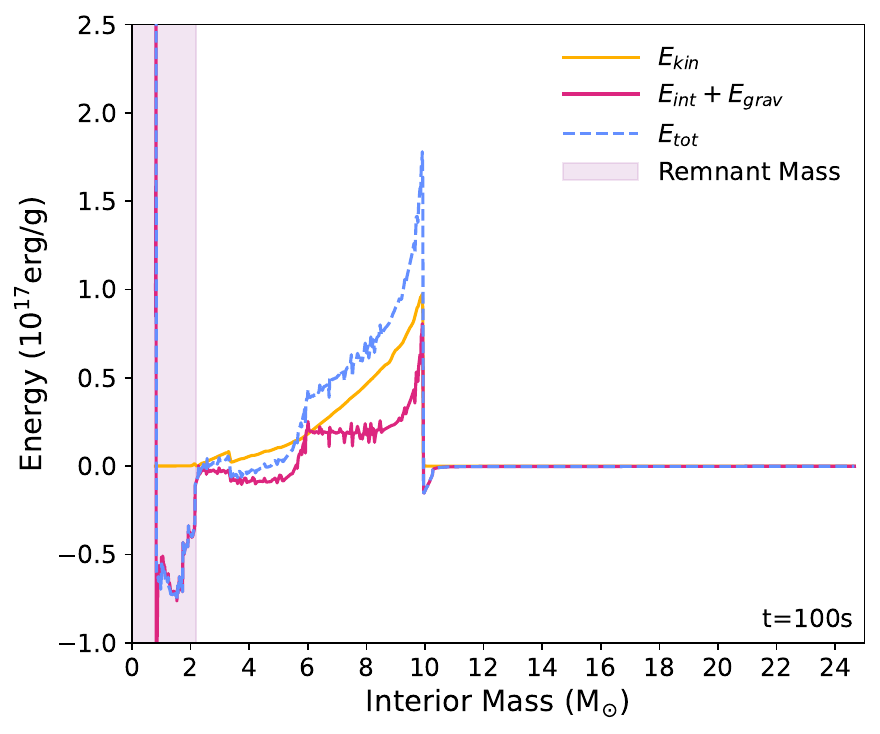}\\
    \includegraphics[width=\columnwidth]{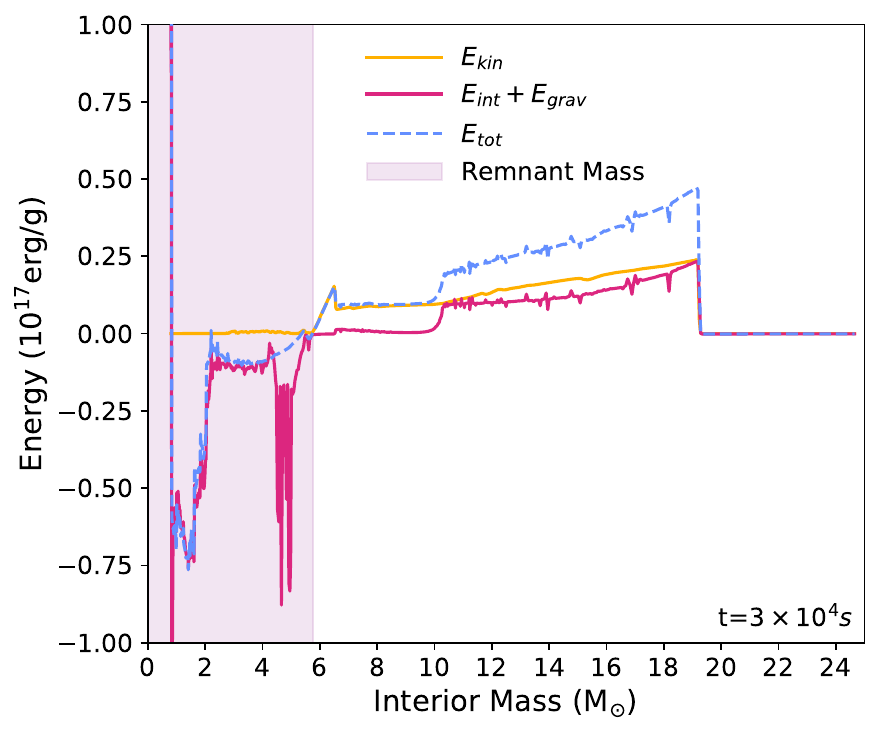}\quad
    \includegraphics[width=\columnwidth]{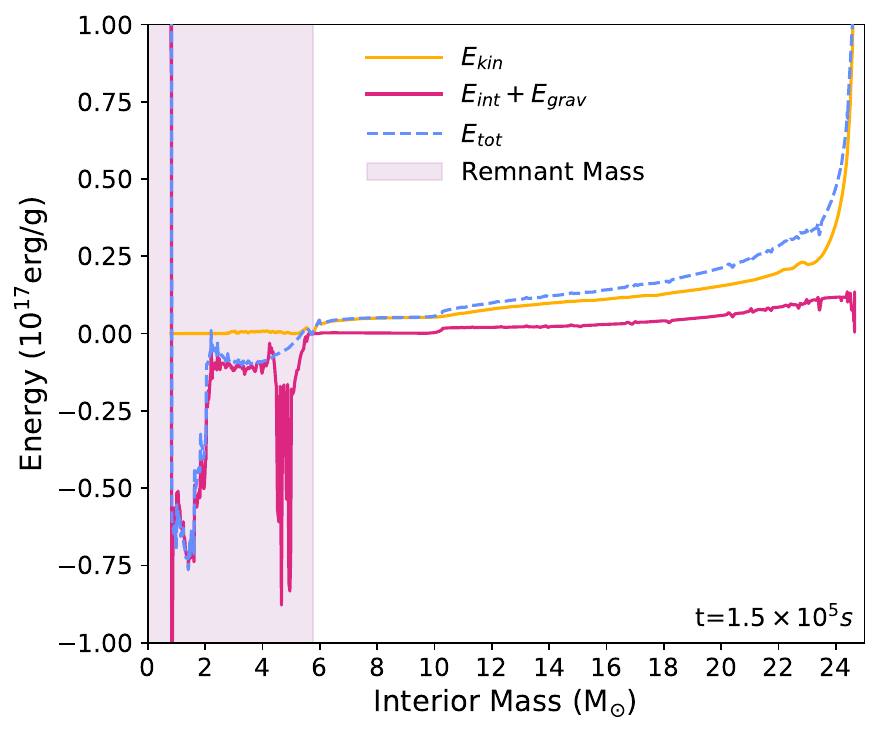}
    \caption{Interior profiles of the kinetic energy (yellow line), the internal energy plus the gravitational energy (purple line), and the total energy (dashed light blue line) at various times during the explosion for a non-rotating stellar progenitor with $\Mzam =25\,\Ms$ with metallicity $\met=-2$. The shaded purple area in the panels represents the region of the star forming the compact remnant at each time. It is completely formed at $t=3\times10^4 \rm s$.}
    \label{fig: SN Energies evolution}
\end{figure*}

The density profile (see Figure \ref{fig: Physical Composition 25Ms progenitor}) shows gradients at mass coordinates which correspond to boundaries between shells of different chemical composition.
The most prominent density gradient is at the mass coordinate of $\sim 10\,\Ms$ and it corresponds to the interface between the He core and the H envelope.
The injection of thermal energy into the iron core results in the formation of a shock wave that moves outward in mass and compresses, heats up, and accelerates the overlying layers.
Once the shock emerges from the iron core, it propagates through the outer layers, where it triggers the explosive nucleosynthesis.
Figure \ref{fig: SN vel} shows the evolution of the velocity profile of the shock wave at different times after the onset of the explosion. 
After $\sim 0.7 \, \rm s$, the shock wave reaches the mass coordinate of $\sim 2.7 \, \Ms$, where the peak temperature of the shock becomes too low to trigger any other additional burning. Therefore, from this mass coordinate outward, the chemical structure sculpted by the hydrostatic evolution will be unaffected by the passage of the shock wave.\par
After $15.5$\, s from the beginning of the explosion, while the shock is still moving through the CO core, some of the innermost layers (i.e. the inner  $\sim 2 \, \Ms$) revert their motion because their velocity is lower than the local escape speed (see the negative values of the orange line in Figure \ref{fig: SN vel}).\par
At $\sim 100\, \rm s$, the shock reaches the He/H interface at a mass coordinate of $\sim 10\,\Ms$, where a strong density gradient (see Figure \ref{fig: Physical Composition 25Ms progenitor}) causes the formation of a reverse shock \citep[see][]{Woosley1995}. From this time onward, the explosion is characterized by a shock wave moving outward in mass and by a reverse shock, which propagates inward in mass, slowing down the previously accelerated material. This effect is apparent from the behavior of the velocity of the shock $1000$ s after the onset of the explosion (red dashed line in Figure \ref{fig: SN vel}), for which we find that the velocity of the outer regions of the shock is smaller than that of the innermost regions of the shock wave.
This happens because the latter have not yet been affected by the reverse shock at this stage of the explosion.\par
Furthermore, Figure \ref{fig: SN vel} shows also that when the shock reaches the He/H interface, the compact remnant increases its mass up to $\sim 2 \, \Ms$, since the layers which had negative velocity when the shock has reached the CO core have now reached zero velocity. 

The fallback process eventually ends $\sim3\times10^4 \, \rm s$ after the onset of the explosion, leaving a compact remnant with a final mass of $5.77 \, \Ms$ (the purple shaded area in Figure \ref{fig: SN vel}).
It is worth noting that such a heavy compact remnant includes also the layers where the explosive nucleosynthesis happened ($\lesssim 2.7\,  \Ms$, i.e. the Si and ONe shells, see Figure \ref{fig: Chemical Composition 25Ms progenitor}), preventing their ejection into the interstellar medium unless some mixing of material from the innermost zones to the outer layers is occurring during the explosion.
Finally, $\sim 1.5\times10^5 \, \rm s$ (i.e. $\sim 1.7$ days) after the beginning of the explosion, the forward shock reaches the surface of the star (shock breakout).\par 
Figure \ref{fig: SN Energies evolution} shows the kinetic energy, the sum of internal and gravitational energy, and the total energy of the star as a function of the mass coordinate, at times $> 15$ seconds after the onset of the explosion. From this figure it is apparent that, at each time, some regions are falling back onto the proto-compact object (i.e. those with negative total energy) and some are being expelled by the shock wave (i.e. those with positive total energy). The last region with negative total energy corresponds to the boundary of the final remnant, that is represented in Figure \ref{fig: SN Energies evolution} with a purple shaded area.\par
\begin{figure}[htb]
    \includegraphics[width=\columnwidth]{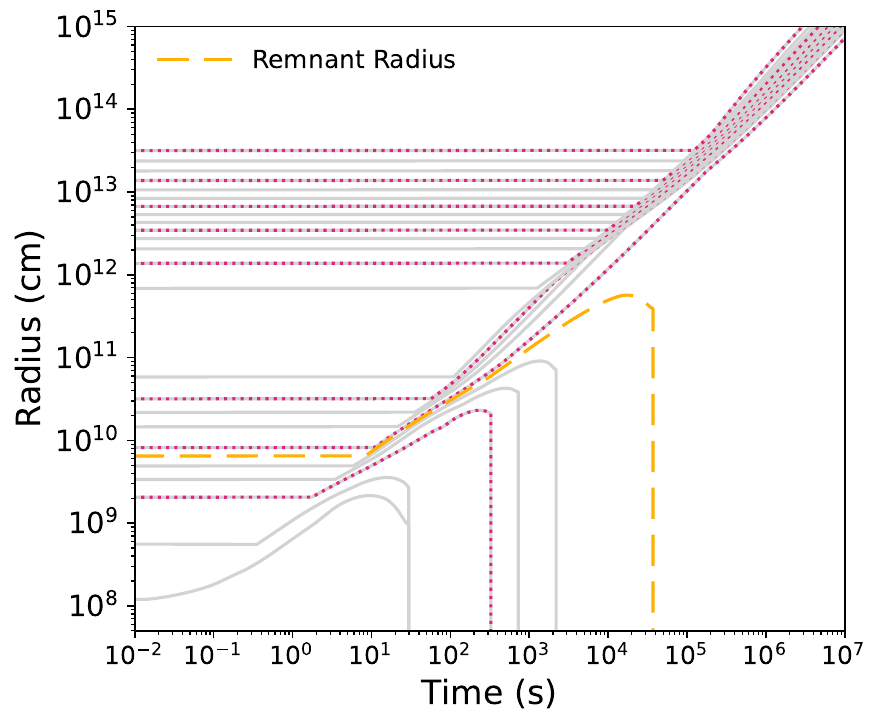}
    \caption{Evolution of the radius of various shells (grey and dotted purple lines) inside the stellar progenitor as a function of time. Each grey (dotted purple) line on top encloses $1\,\Ms$ ($3\, \Ms$) more than the grey (dotted purple) line immediately below.  We also show the evolution of the radius of the final compact remnant (yellow dashed line).}
    \label{fig: Succesfully SN case}
\end{figure}

We present the whole process in Figure \ref{fig: Succesfully SN case}, which shows the evolution of the radius of various mass shells (grey, in step of  $1 \Ms$, and dotted red lines, in step of $3 \Ms$) inside the stellar progenitor as a function of time. From this figure we can distinguish the layers of the star that fall back onto the compact object, i.e., those that form the final remnant (represented by the dashed yellow line), and those that are successfully ejected at the shock breakout. Roughly at $\sim10\,$s we find that the radius of the mass shells corresponding to the most internal $2\,\Ms$ of the star (the first two grey lines) revert their trajectory, start to fall back onto the proto-compact object, and they reach the latter $\sim100\,$s after the onset of the explosion.\par
Figure \ref{fig: Succesfully SN case} also shows when the fallback ends. The last shell that falls back onto the compact object reverts its trajectory roughly $\sim 10^4$ s after the onset of the explosion and it reaches the compact remnant at $t=3\times 10^4$ s (consistent with our results reported in Figure \ref{fig: SN vel} and in the bottom left panel of Figure \ref{fig: SN Energies evolution}). For $t\geq 3\times 10^4$ s, such mass shell has zero velocity forming the final remnant with mass $5.77\,\Ms$. In Figure \ref{fig: Succesfully SN case} all the layers that eventually form the final remnant are shown to instantaneously drop onto the compact remnant but this is only a numerical effect, since once they revert their trajectory, falling back onto the BH, we do not follow anymore their evolution, i.e. we do not observe the free fall trajectories.

In contrast, the outer region is eventually ejected by the shock wave at the shock breakout, and it keeps moving outward in the interstellar medium with a radial velocity that grows linearly with time.\par

\subsection{Direct collapse} \label{sec: failed-SN run}
\begin{figure}[htb]
    \centering
    \includegraphics[width=\columnwidth]{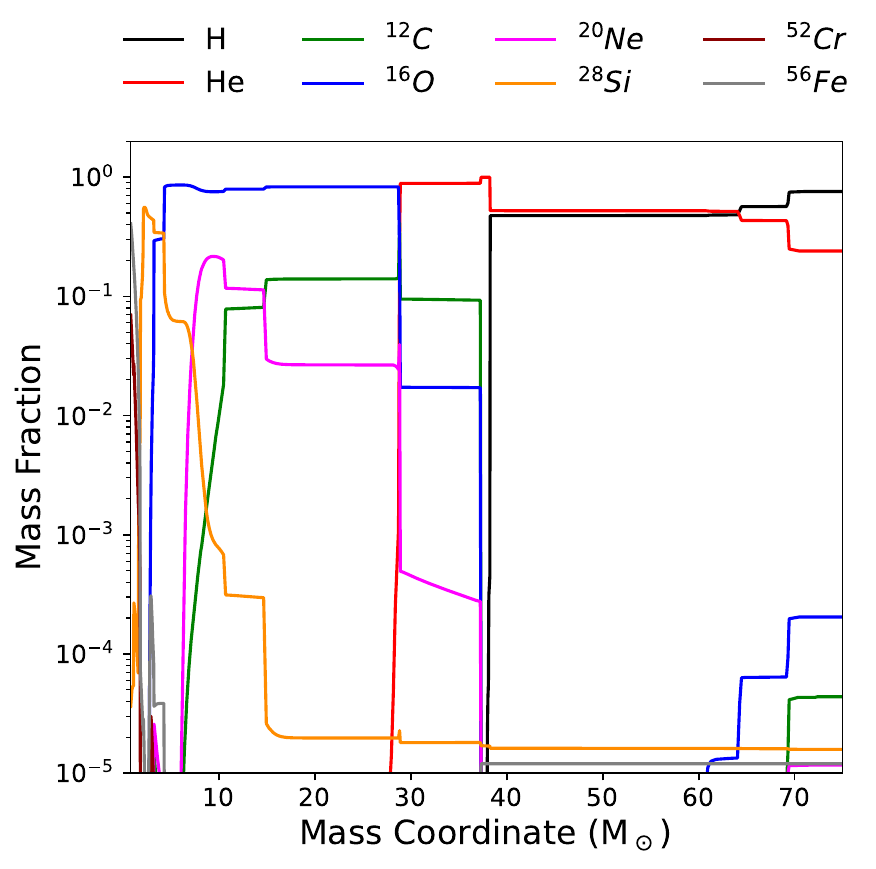}       
    \caption{Chemical composition as a function of the interior mass coordinate of a non-rotating model with $\Mzam=80\,\Ms$ and $\met=-2$ at the pre-SN stage. We report the abundances of H (long dash-double dotted black line), He (double dash-dotted red line), $^{12}C$ (long dashed green line), $^{16}O$ (short dashed blue line), $^{20}Ne$ (dashed magenta line), $^{28}Si$ (short dash-double dotted orange line), $^{52}Cr$ (solid dark red line) and $^{56}Fe$ (long dash-dotted gray line).}
    \label{fig: Chemical Composition 80Ms progenitor}
\end{figure}

\begin{figure}[htb]
    \centering
    \includegraphics[width=\columnwidth]{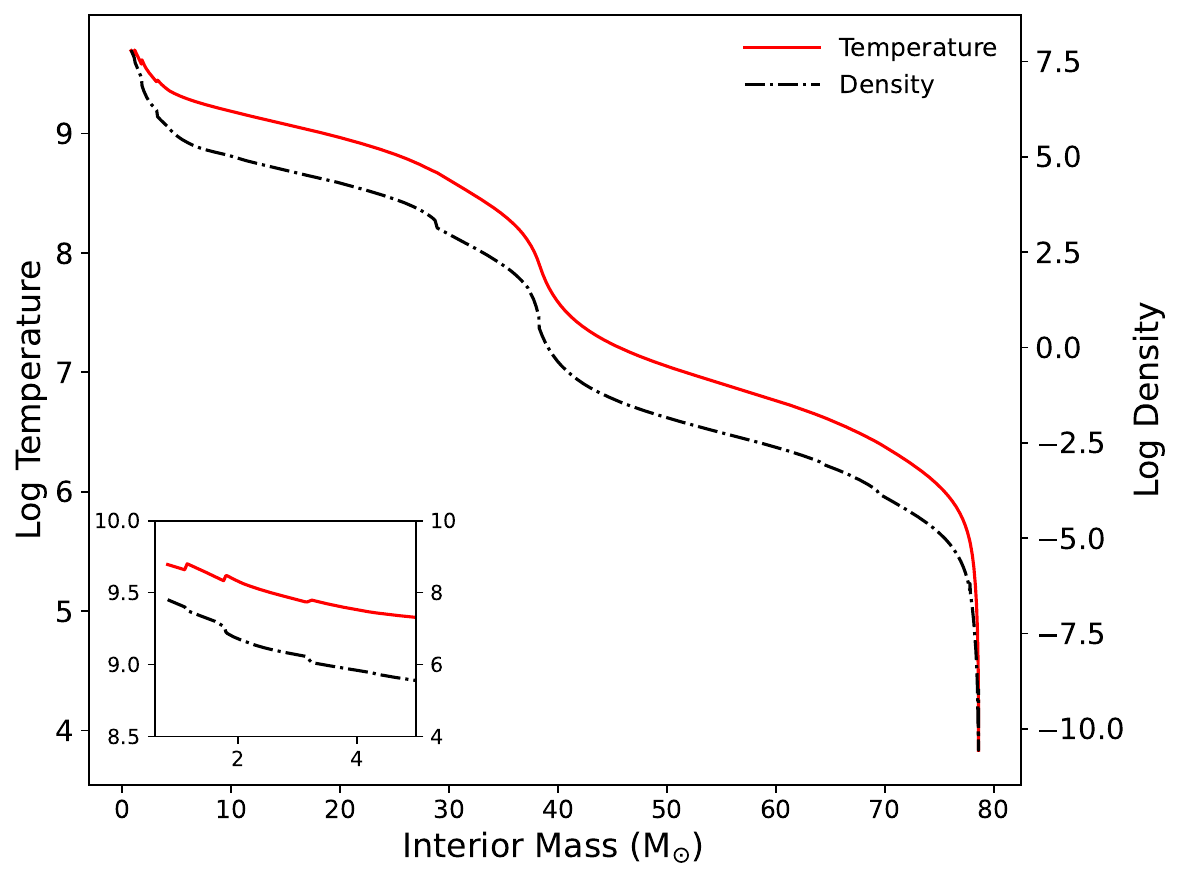}
    \caption{Temperature profile (red line) and density profile (dash-dotted black line) at the pre-SN stage as function of the mass coordinate of a non-rotating stellar progenitor with $\Mzam=80\,\Ms$ and $\met=-2$. A zoom-in on the innermost region of the core, up to $5\,\Ms$, is highlighted in the lower right box .}
    \label{fig: Physical Composition 80Ms progenitor}
\end{figure}

\begin{figure}
    \centering
    \includegraphics[width=\columnwidth]{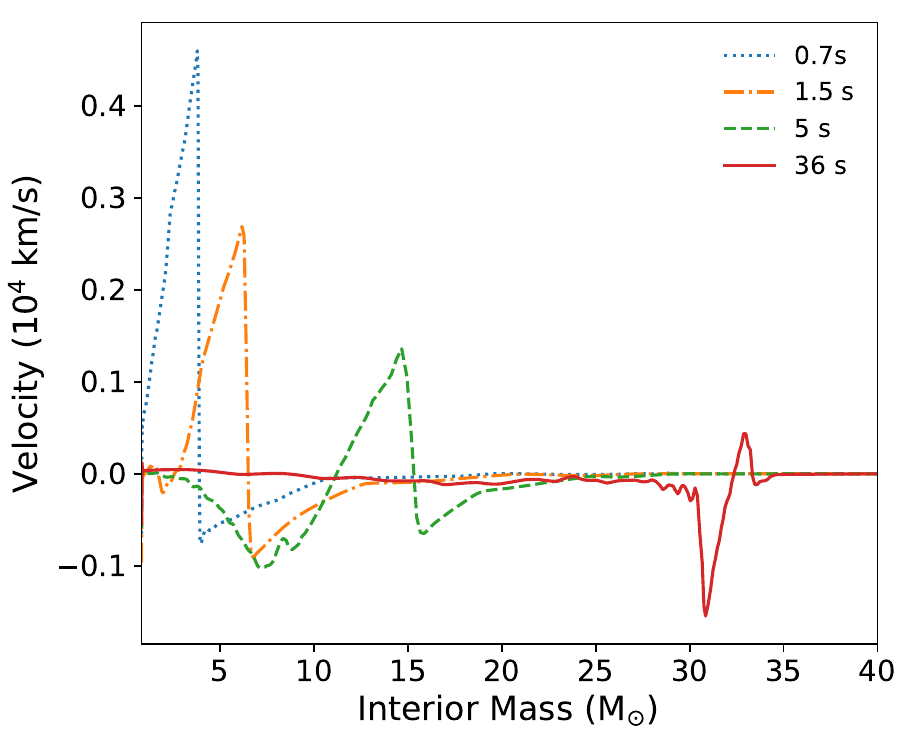}
    \caption{Same of Figure \ref{fig: SN vel} but for a directly collapsing progenitor. The velocity profile is reported at different crucial times during the explosion: when the explosive nucleosynthesis stops (dotted blue line), when the layers behind the shock start to fallback onto the compact remnant (dash-dotted orange line), when the shock enters the CO core (dashed green line) and when the velocity of the shock becomes negligible (solid red line).}
    \label{fig: BH vel}
\end{figure}
The progenitor model we adopt to describe a typical failed explosion which undergoes a direct collapse (DC) to BH in our simulations is a non-rotating $80 \, \Ms$-star with metallicity $\met=-2$. 
Figure \ref{fig: Chemical Composition 80Ms progenitor} shows the chemical structure of the star at the pre-SN stage. The structure is characterized by an extended H-rich envelope, a He core mass of $38.3\, \Ms$ and a CO core mass of $28.9 \, \Ms$. The ONe shell extends between mass coordinate $4.33 \, \Ms$ and $14.9 \, \Ms$ while the Si shell extends from $1.76 \, \Ms$ to $4.33 \, \Ms$. The mass of the iron core is $1.76 \Ms$.\par 
Figure \ref{fig: Physical Composition 80Ms progenitor} shows the temperature and the density profile of the progenitor as function of the mass coordinate. From this figure it is apparent that the density profile presents various steep density gradients, each one corresponding to the boundaries  between shells of different chemical composition.
As in the previous case of a successful explosion, this star loses a negligible amount of mass during its life because of its initial low metallicity and it reaches the pre-SN phase as a RSG, having an effective temperature of $6.3\times10^3 \, \rm K$.\par
In Figure \ref{fig: BH vel}, we show the velocity profile of the explosion as a function of the mass coordinate at different times after the onset of the explosion. Figure \ref{fig: BH vel} shows that the velocity of the shock progressively decreases with time.
The regions behind the shock wave begin reverting their trajectory $\sim 1.5$ s after the onset of the explosion, since they have exhausted their initial kinetic energy. Thus, they start falling back onto the proto-compact object with negative radial velocity. As the shock moves at larger mass coordinates, more and more internal layers exhaust their kinetic energy, and fall back onto the core, until the velocity of the shock becomes almost zero at $\sim 36 \, \rm s$ after the onset of the explosion. Therefore, in this case, the shock wave does not have enough energy to eject the mantle of the star.\par
\begin{figure}[htb!]
    \includegraphics[width=\columnwidth]{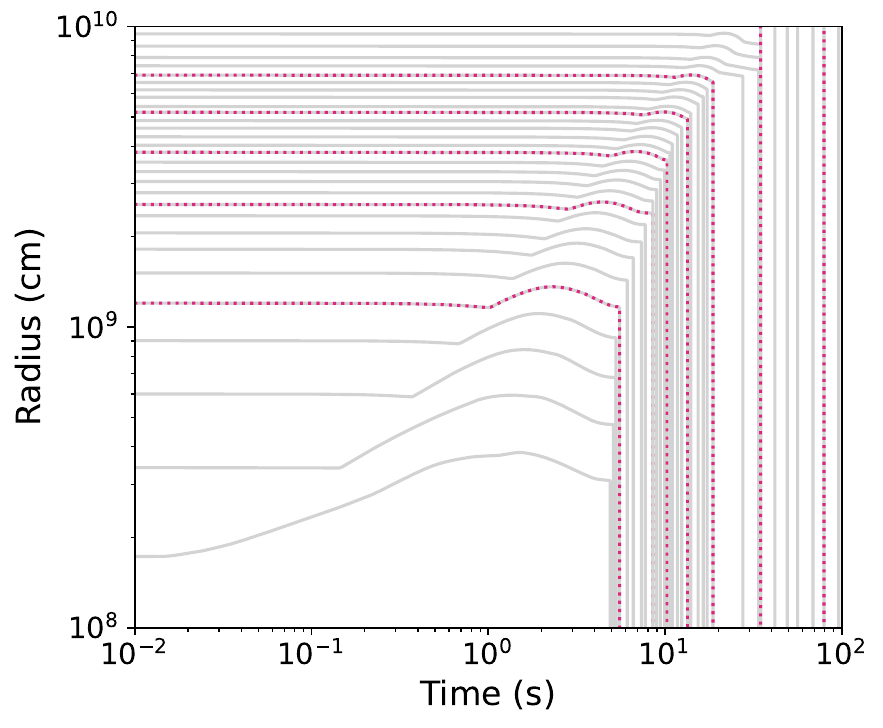}
    \caption{Same of Figure \ref{fig: Succesfully SN case} but for the most central region of the progenitor with $\Mzam=80\,\Ms$ and $\met=-2$, where we can better appreciate the fallback of the layers right after being accelerated by the shock wave. We show the behavior of the shell enclosing different values of mass, of $1\,\Ms$ (each solid light grey lines) and $5\,\Ms$ (each dotted purple line), respectively.}
    \label{fig: Direct Collapse case}
\end{figure}

Figure \ref{fig: Direct Collapse case} shows the dynamics of the innermost layers of the stellar progenitor. Such layers undergo a brief acceleration phase when the shock wave reaches them. However, such acceleration is not sufficient to unbound these regions from the star. In Figure \ref{fig: Direct Collapse case} we notice the same behaviour we presented in Figure \ref{fig: BH vel}: already $\sim 1.5$ s after the onset of the explosion, the innermost layers of the star revert their trajectory, falling back onto the proto-compact object, reaching the latter only a few seconds later, and becoming part of the final remnant at $\sim 5$ s. Furthermore, the acceleration decreases in the outer layers, which start to collapse again immediately after the passage of the shock wave. Therefore, the latter cannot reach the surface of the star and the entire progenitor directly collapses into a heavy BH.

\subsection{Remnants BHs from non-rotating progenitors} \label{sec: non-rot progenitors}

\begin{figure}[htb!]
    \centering
    \includegraphics[width=\columnwidth]{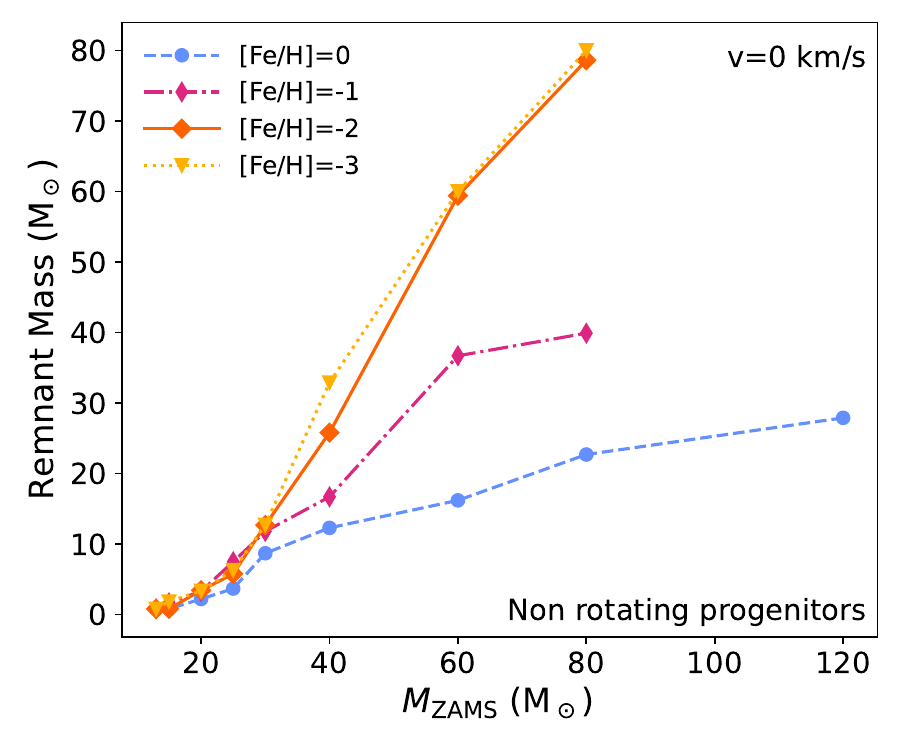}
    \caption{Remnant masses as a function of $\Mzam$ for the progenitors with initial metallicity $\met=0$ (dashed light blue line), $\met=-1$ (dash-dotted violet line), $\met=-2$ (orange solid line), and $\met=-3$ (dotted yellow line). We do not simulate the explosion of the stellar progenitors with $\Mzam=120\,\Ms$ at subsolar metallicities, since they are unstable against pair production.}
    \label{fig: plot mrem NR}
\end{figure}
\begin{figure*}[htb!]
    \centering
    \includegraphics[width=\textwidth]{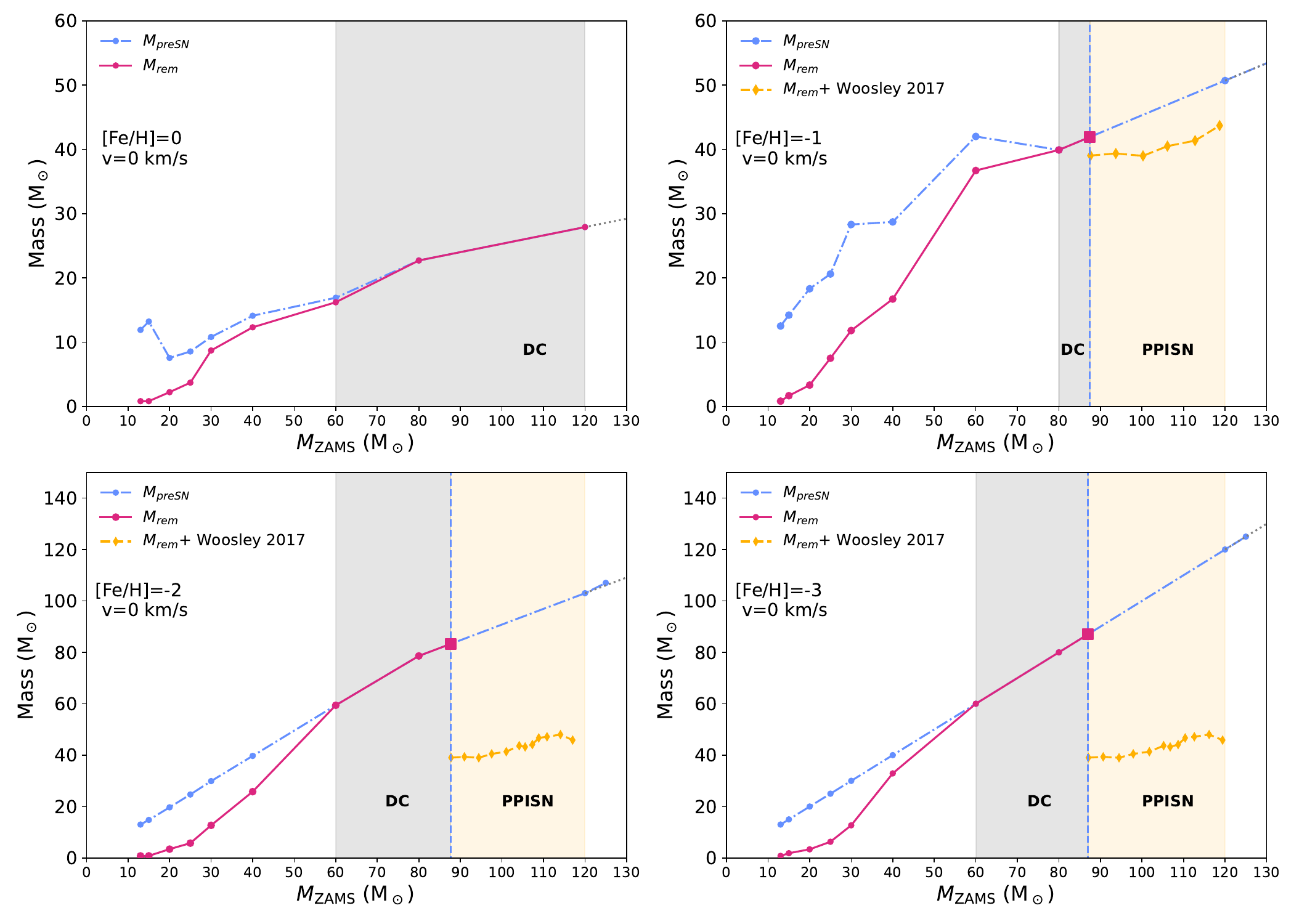}
    \caption{ Pre-SN mass (dash-dotted light blue line), pre-SN mass extrapolated for $\Mzam > 120 \Ms$ (dotted grey line), remnant mass (solid violet line), and the remnant mass for the stars which develop pair instabilities (dashed yellow line), as a function of $\Mzam$. The red points represent the stellar progenitors of the FRANEC grid, while the larger red squares correspond to the last progenitor stable against pair production, which has been obtained through interpolation (see Section \ref{sec: PI regime} and Table \ref{Tab: Interpolation results}). The shaded grey area represents the region of DC, the shaded yellow area shows where stars undergo PPISNe. The vertical dashed lines represent the threshold values above which stars enter the PI regions.}
    \label{fig: Mpre-SN and Mrem vs Mini NR case}
\end{figure*}

Figure \ref{fig: plot mrem NR} shows the remnant masses as a function of $\Mzam$, for non-rotating progenitors and for various initial metallicities. From Figure \ref{fig: plot mrem NR} it is apparent that the remnant mass increases as a function of $\Mzam$ for all the considered metallicities. Furthermore, the remnant mass is larger for progenitors at lower metallicities, and does not depend significantly on metallicity for progenitors with $\Mzam \lesssim 25\,\Ms$. \par
The remnant mass curves follow closely the behavior of the CO core mass (see the bottom-left panel of Figure \ref{fig: plot co and mpresn vs mzams}).\par
This happens because the remnant mass depends on the binding energy of the star at the pre-SN stage, and that comes mainly from the CO core. Moreover, larger CO core masses imply lower $\rm ^{12}C$ mass fractions at core-He depletion, i.e. a less efficient C-shell burning, which, in turn, causes a higher contraction of the CO core resulting in a more bound structure. \par
Figure \ref{fig: Mpre-SN and Mrem vs Mini NR case} shows the mass of the progenitor star at the pre-SN stage and the mass of the compact remnant as a function of the initial mass of the star, for various metallicities. In our simulations, we assume that DC occurs if the fraction of the progenitor's envelope ejected by the shock wave is less than 10\% of the pre-SN mass. For example, the progenitor at solar metallicity with $\Mzam=60\,\Ms$ has a pre-SN mass of $16.9\,\Ms$ and forms a remnant of $16.2\,\Ms$, ejecting only $0.7\,\Ms$ of H envelope, thus we consider it a DC.
Following this approach, we find that stars with $\Mzam \gtrsim 60\,M_\odot$ do not explode and collapse directly to a BH for the initial metallicities  $\met=0$, $\met=-2$, $\met=-3$. For $\met=-1$, the threshold for DC is $\Mzam= 80\,\Ms$. 
This happens because the progenitor with $\Mzam=60\,\Ms$ ends its life as a RSG. Thus, the shock wave ejects part of the H envelope since this is very loosely bound to the star (see \cite{Fernandez2018}). This happens only for this value of metallicity, because at $\met=0$ the star with $\Mzam=60\,\Ms$ has already lost all its H envelope at the pre-SN stage, while at $\met=-2$ and $\met=-3$ stellar winds are quenched, i.e. all the H envelope is retained and the star is much more compact. \par
We do not account for additional mass loss effects associated with DC (e.g., from neutrino emissions, as proposed by \citealt{Nadezhin_1980, Piro_2013}). In our simulations, all massive stars that end their lives through DC are compact BSG stars. Therefore, such effects are negligible (see \citealt{Lovegrove2013, Fernandez2018}), as they primarily affect more extended stars like RSGs. However, in our simulations such stars naturally lose their H envelope as a consequence of the shockwave.\par
The BH masses at the threshold for DC depend on the final mass of the progenitor stars at the pre-SN stage that, in turn, depends on the mass-loss history during the pre-SN evolution. The mass-loss rate decreases significantly at low metallicity, where stellar winds are quenched.
Therefore, the models with initial metallicity lower than $\met=-1$ evolve approximately at constant mass. As a consequence, the BH masses correspond to the threshold in $\Mzam$ for DC (see Table \ref{Tab: Interpolation results}) .

Above $\Mzam^{\rm PPISN}$ we compute the remnant masses taking into account the results of W17 (see, e.g., its Table 1) by interpolating on the CO core mass.

The results are reported in Figure \ref{fig: Mpre-SN and Mrem vs Mini NR case} as yellow diamond points and are connected through the dashed yellow lines.
From this Figure it is apparent that at $\met=-1$ the onset of PPISNe does not affect significantly the maximum BH mass that forms. This happens because - at the pre-SN stage - stars have already lost a significant part of their H envelope, thus the mass loss caused by PPISNe does not affect significantly the final BH mass. In contrast, for $\met=-2,\,-3$, stars evolve at almost constant mass and end their life with heavy H envelopes. Therefore, the expulsion of the latter at the onset of PPISN significantly affects the final BH mass. This effect is apparent in the bottom-row panels of Figure \ref{fig: Mpre-SN and Mrem vs Mini NR case}, where we see a steep variation of the final remnant mass at the onset of PPISNe.\par

\subsection{Remnants BHs from rotating progenitors} \label{sec: rot progenitors}

\begin{figure}[htb!]
    \centering
    \includegraphics[width=\columnwidth]{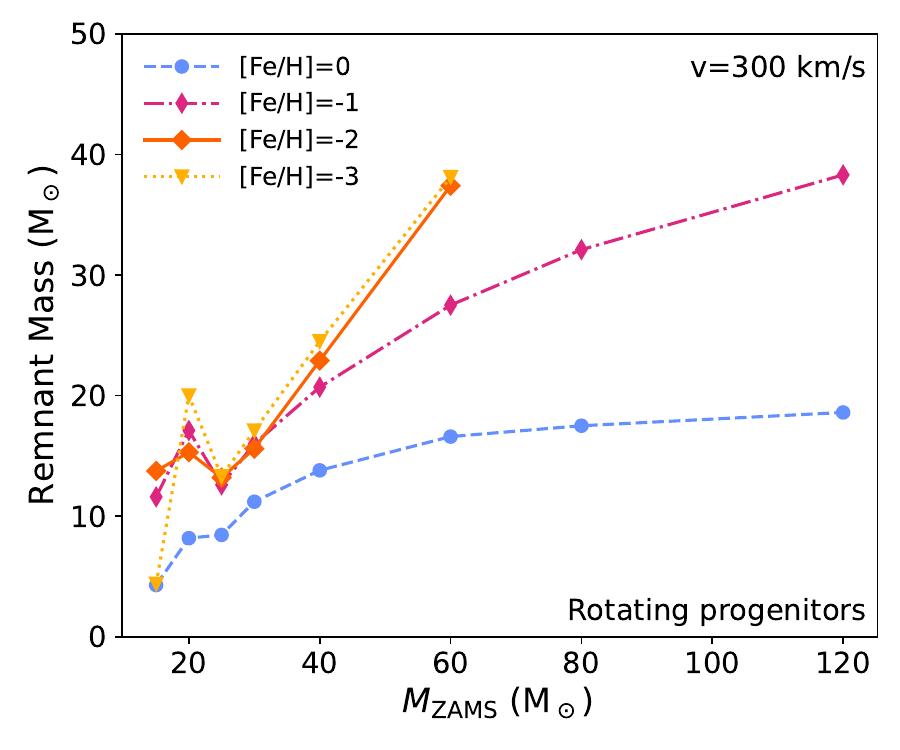}
    \caption{Same of Figure \ref{fig: plot mrem NR} but for rotating progenitors.  We do not simulate the explosion of the stellar progenitors with $\Mzam=80\,\Ms$ and $120\,\Ms$ at $\met=-2,\,-3$, since they are unstable against pair production.}
    \label{fig: plot mrem R}
\end{figure}
\begin{figure*}[htb!]
    \centering
    \includegraphics[width=\textwidth]{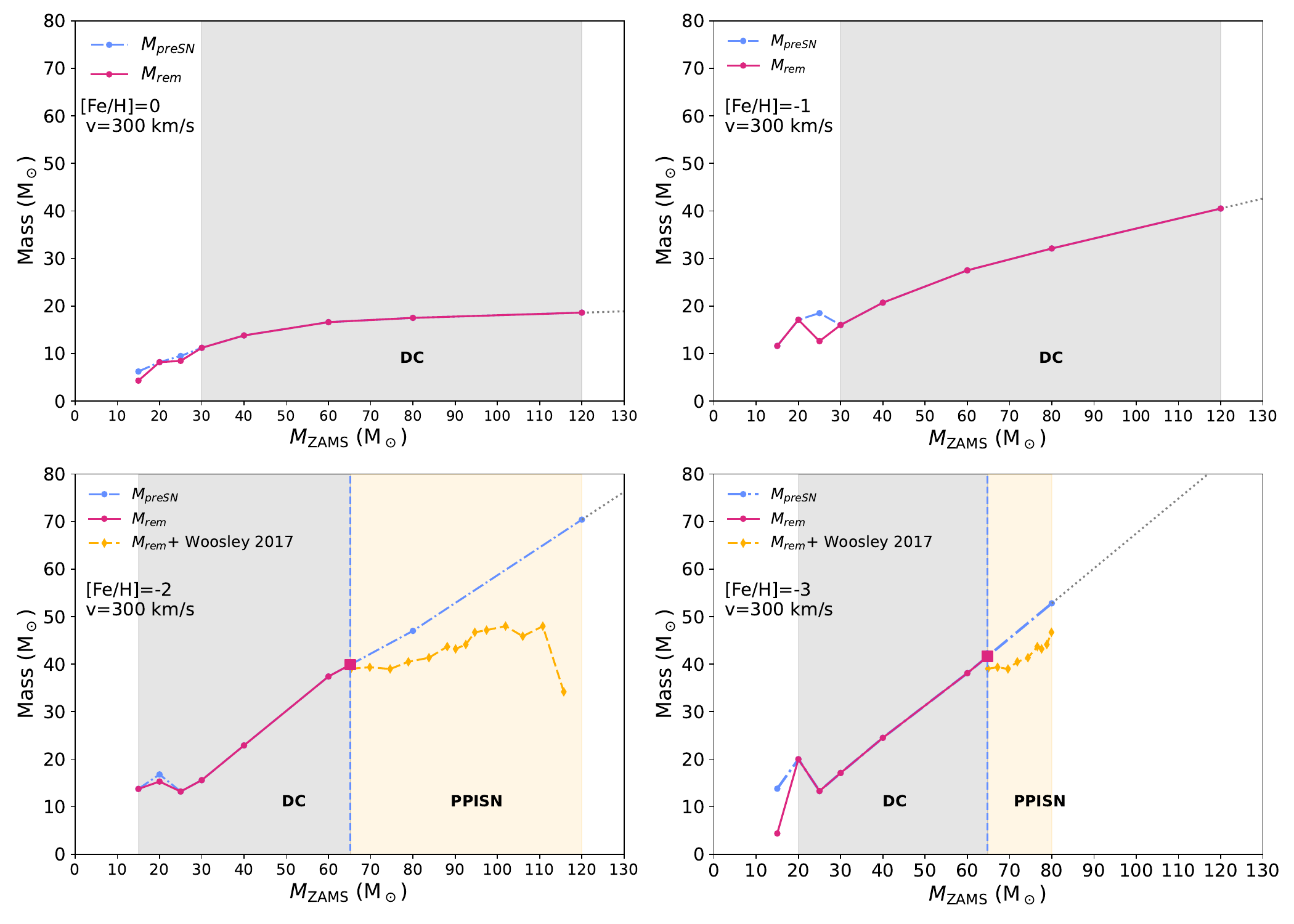}
    \caption{Same as Figure \ref{fig: Mpre-SN and Mrem vs Mini NR case} but for rotating progenitors.}
    \label{fig: Mpre-SN and Mrem vs Mini R case}
\end{figure*}

Figure \ref{fig: plot mrem R} shows the remnant masses as a function of $\Mzam$, for rotating progenitors and for various initial metallicities. 
From Figure \ref{fig: plot mrem R} it is apparent that, at solar metallicity, the remnant mass does not increase significantly with the initial progenitor mass. In contrast, for $\met = -1,\,-2,\,-3$, the remnant mass curve has a local minimum at $25 \,\Ms$, which occurs when the stellar winds are strong enough to completely remove the H envelope. Furthermore, as in the case of the non-rotating progenitors, for $\Mzam \gtrsim25\,\Ms$ the remnant masses are larger at lower metallicities, and follow the behaviour of the CO core mass at the pre-SN stage (see also Section \ref{sec: non-rot progenitors}).\par
Figure \ref{fig: Mpre-SN and Mrem vs Mini R case} shows how the remnant mass and the mass of the star in the pre-SN phase depend on $\Mzam$, for rotating models with different initial metallicities.
We find that most of the rotating progenitors 
directly collapse to BHs.
This happens because rotating stars end their life as He-stars, i.e. they end their life with more bound structures at the pre-SN stage with respect to non-rotating progenitors, thus they are more prone to failed SN explosions and to form more massive BHs.
Specifically, stars with  $\Mzam \gtrsim 30\,\Ms$ collapse directly to a BH, whereas the fate of stars with $\Mzam \lesssim 30 \,\Ms$ depends on the initial metallicity.\par
At solar metallicity, the threshold value of $\Mzam$ for the DC decreases down to $\sim20\,\Ms$, since only the progenitor with $\Mzam=15\,\Ms$ evolves through a successful SN explosion. At $\met=-1$, the only exploding progenitor is the one with $\Mzam=25 \, \Ms$, and it ejects all its H envelope, while the less massive progenitors directly collapse to BHs. Although this occurs only for one progenitor in the simulation set, the non-monotonic behaviour for the explodability is not apparent for non-rotating progenitors (see Figure \ref{fig: Mpre-SN and Mrem vs Mini NR case}). At $\met=-2$, all the progenitors evolve through DC. The progenitor with $\Mzam=20\,\Ms$ ejects some material but it is still consistent with our definition of DC, thus the threshold value of $\Mzam$ for the DC is $\sim15\,\Ms$. The stars at $\met=-3$ behave similarly to solar metallicity models, and the only exploding progenitor is the one with $\Mzam=15\,\Ms$, thus the threshold of $\Mzam$ for the DC is at $\sim20\,\Ms$.\par
For rotating progenitors, mass loss is enhanced because rotation forces the models to become red giants and therefore to approach their Eddington luminosity during their redward excursion in the HR diagram (see LC18). When this happens, a substantial amount of mass is lost and most of the stellar models loose all their H rich envelope and become Wolf-Rayet stars. Therefore, progenitors with the same $\Mzam$ and metallicity retain less mass at the pre-SN stage than non-rotating stars.
This is apparent especially at lower metallicities, where stellar winds for non-rotating progenitors are quenched. This effect is not significant for progenitors with $\Mzam\leq 30\,\Ms$, since such stars have weak stellar winds. As a consequence, we obtain less massive BHs from rotating stellar progenitors than from non-rotating ones.\par
The BH masses corresponding to the threshold in $\Mzam$ for the DC are $\sim 8\,\Ms$, $\sim 16\,\Ms$, $\sim 14\,\Ms$ and $\sim 20\,\Ms$ for metallicities $\met=0$, $\met=-1$, $\met=-2$, $\met=-3$, respectively. The threshold CO core mass for PPISNe, $\rm M_{CO}^{PPISN}\sim 33\,\Ms$ are reported in Table \ref{Tab: Interpolation results}.

To estimate the remnant masses of the models with $\Mzam\geq\Mzam^{\rm PPISN}$ we apply the results of W17, as described in Section \ref{sec: non-rot progenitors}. The results of such interpolations are the diamond yellow points, which are connected through the dashed yellow lines. From Figure \ref{fig: Mpre-SN and Mrem vs Mini R case}, it is apparent that, for rotating stellar progenitors, there is no major change in the BH mass due to the onset of PPISNe. 
This happens because the rotating stars end their life as naked He stars, thus they have lost their H envelope. Therefore, the rotating stars only lose a small fraction of their He core at the onset of the PPISNe, because of the pulsational phase, and then collapse to to BHs.
\section{Discussion}
\label{sec: Discussion}

\subsection{The maximum BH mass}\label{section: Max BH mass}

\begin{figure}[htb!]
    \centering
    \includegraphics[width=\columnwidth]{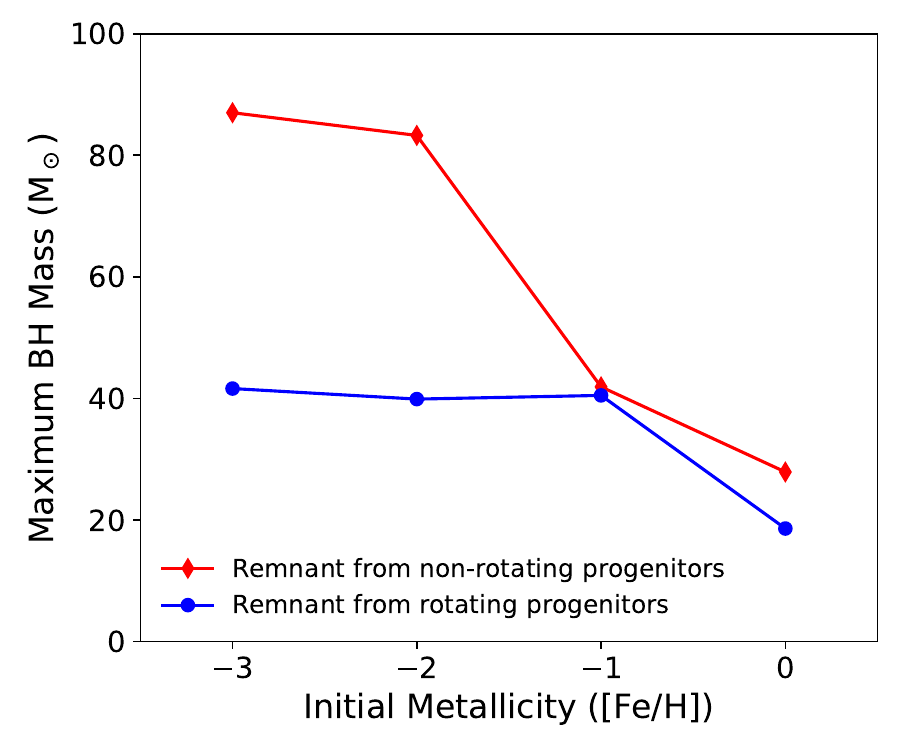}
    \caption{Maximum BH mass as a function of the initial metallicity for the non-rotating stellar progenitors (red line) and the rotating stellar progenitors (blue line) for the non-rotating progenitors.}
    \label{fig: Max BH mass}
\end{figure}

Figure \ref{fig: Max BH mass} shows the maximum BH mass resulting from our simulations as a function of the initial stellar metallicity, and assuming rotating and non rotating single star models. It illustrates the impact of the initial metallicity and  angular rotation of the stellar progenitor on the final remnant mass we obtain from single stellar evolution. \par
For $\met=0$ and $\met= -1$, stellar winds play a crucial role as they remove a significant fraction of mass of the progenitor stars before they DC to BHs. From our simulations we find that the maximum BH mass we form is $\sim 27.9\,\Ms$ and $\sim 41.9\,\Ms$, for $\met=0$ and $\met=-1$ in the non-rotating case, and $\sim 18.6\,\Ms$ and $\sim 40.5\,\Ms$, for $\met=0$ and $\met=-1$ in the rotating case. All these BHs come from the collapse of stellar progenitors with $\Mzam=120\,\Ms$.
Therefore, we find that angular rotation slightly enhances the effects of the stellar winds at solar metallicity, while at $\met=-1$ the effect of rotation on the final remnant masses is negligible. 
For the cases $\met=-2, -3$ the heaviest BHs from non-rotating stellar progenitors have mass $\sim 83.3\,\Ms$ and $\sim 87.0\,\Ms$ respectively, and they both correspond to stellar progenitors with $\Mzam\simeq 87\,\Ms$ (as reported in Table \ref{Tab: Interpolation results}). In the rotating case, the maximum BH mass is $\sim 39.9 \, \Ms$ and $\sim 41.6\,\Ms$ for $\met=-2$ and $\met=-3$, respectively, and they correspond to stellar progenitors with $\Mzam\simeq65\,\Ms$ (see Table \ref{Tab: Interpolation results}). Therefore, in the rotating scenario, the initial angular rotation enhances mass loss and induces the formation of larger convective regions and therefore larger CO cores (see LC18, \citealt{marassi_2019}, and \citealt{mapelli_impact_2020}). As a consequence, stars are less massive at the pre-SN stage and enter the PPISNe regime at lower initial masses compared to non-rotating progenitors. Such a combination of effects limits significantly the value of the maximum BH mass.

\subsection{Implications for the PI mass gap}

\begin{figure}[htb]
    \centering
    \includegraphics[width=\columnwidth]{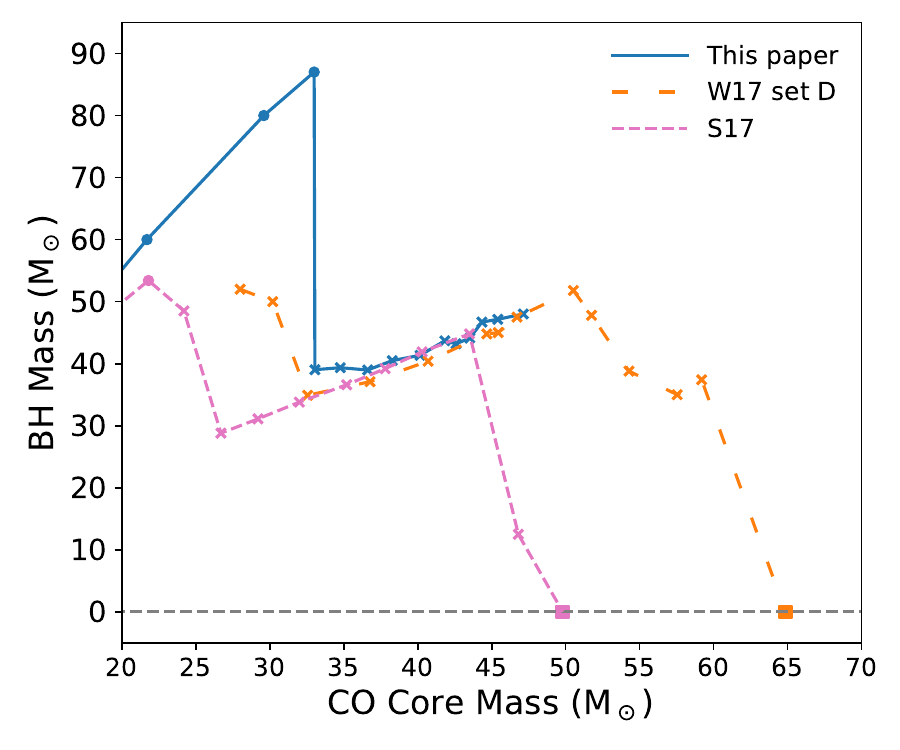}
    \caption{BH mass spectrum as a function of the CO core mass from various authors. Solid blue line: this work, non-rotating progenitors at $\met =-3$; long dashed orange line: set D of W17 (note that these models are obtained from stellar tracks computed at $Z=0.1\,\Zs$ for which the mass loss has been  inhibited, making these models distinct from those we used to compute the final remnant masses.); dashed pink line: models from \cite{Spera_2017} with $Z=1.3\times 10^{-2}\,\Zs$ (S17). 
    Circles (crosses) represent the models that are stable (unstable) against pair production, and squares are the stars disrupted by PISNe.} 
    \label{fig: PPI threshold}
\end{figure}
Figure \ref{fig: Max BH mass} shows that we can form heavy BHs that lie in the upper mass gap predicted by PISN models, i.e. a dearth of BHs with masses in the range $60\, \textrm{--}\, 120 \Ms$ (e.g., \citealt{Woosley2017, Spera_2017}).
This is because we can form BHs in the upper mass gap because the FRANEC progenitors enter the pair-instability at different initial progenitor masses than those of other authors (e.g. \citealt{Woosley2017, Spera_2017, Farmer2019, Costa2021}).


Given our choice of the minimum CO core pass for the onset of PPISN of $33\,\Ms$ (see Section \ref{sec: PI regime}), we find that, for the non-rotating models at $\met=-3$, the $\Mzam = 80\, \Ms$ progenitor which ends its life with a CO core mass of $30 \,\Ms$ is stable against pair production. Conversely, the $\Mzam=120\,\Ms$ has a CO core of $\sim 50\,\Ms$ and develops dynamical instabilities. 
To pin down the uncertainty on the maximum BH mass formed we would need to increase the resolution of our grid of progenitors between $\Mzam=80\,\Ms$ and $\Mzam=120\,\Ms$.\par 
To compare our results with other studies, we show in Figure \ref{fig: PPI threshold} the BH mass as a function of the pre-SN CO core mass of the progenitor star as obtained in this work (LC18 non-rotating progenitors at $\met=-3$), W17 (set D of its Table 2, a set of tracks evolved from the ZAMS with metallicity of $Z=0.1\,\Zs$ taking into account also the hydrogen envelope and for which they assume zero mass loss), \cite{Spera_2017} (hereinafter S17) for metallicity of $Z=1.3\times 10^{-2}\,\Zs$.\par
To compare the BH mass spectra, we choose the stellar tracks with zero initial angular velocities, to exclude the effects of chemical mixing and mass loss enhancement, and with the lowest initial metallicities, to exclude degeneracies caused by different stellar-wind models.

Figure \ref{fig: PPI threshold} shows that the first progenitors that evolve through the pulsational pair-instability phase (marked with crosses) are those of S17, with a CO core mass of $\sim 24\,\Ms$. This happens because, to enter the PPISNe regime, S17 adopted the He core mass criterion proposed in W17 (He core mass of $\sim 32\,\Ms$), which corresponds to CO core masses of $\sim 24\,\Ms$ when used in combination with the S17 stellar evolution tracks from the PARSEC code (\citealt{Bressan2012}).\par
The tracks of W17 enter the PPISNe regime when the CO core mass is $\sim 28\,\Ms$. It is important to note that these stellar tracks were evolved suppressing the mass loss phenomena of a set of stars with initial metallicity of $Z=0.1\,\Zs$, and that they were evolved from the ZAMS to the end of the PPISN regime. Thus, they represent a distinct set from the naked helium cores we used in Section \ref{sec: non-rot progenitors} and \ref{sec: rot progenitors} to infer the final remnant masses of the stars undergoing PPISNe. \par
Finally, our stellar progenitors become unstable against pair-instability when the CO core mass is $\gtrsim 33 \,\Ms$.
From Figure \ref{fig: PPI threshold} it is apparent that the BH maximum mass ranges from $\sim 87\,\Ms$ (this work) to the minimum value of $\sim 53\,\Ms$ (S17), resulting in a discrepancy of $\sim 38\,\Ms$ in the maximum mass of the BHs obtained by different authors from the evolution of an isolated massive star. The discrepancy exists because different authors have different thresholds in the CO core mass for entering the PPISNe regime (see also \citealt{Farmer2019}, hereinafter F19). For instance, if we had used the value of $\rm M_{CO}^{PPISN}$ from F19 (e.g., $\rm M_{ CO,F19}^{PPISN} =41\,\Ms$) we would have ended up with BHs more massive than $\sim 100\,\Ms$, i.e. in the regime of intermediate-mass BHs.\par

\begin{figure}[htb]
    \centering
    \includegraphics[width=\columnwidth]{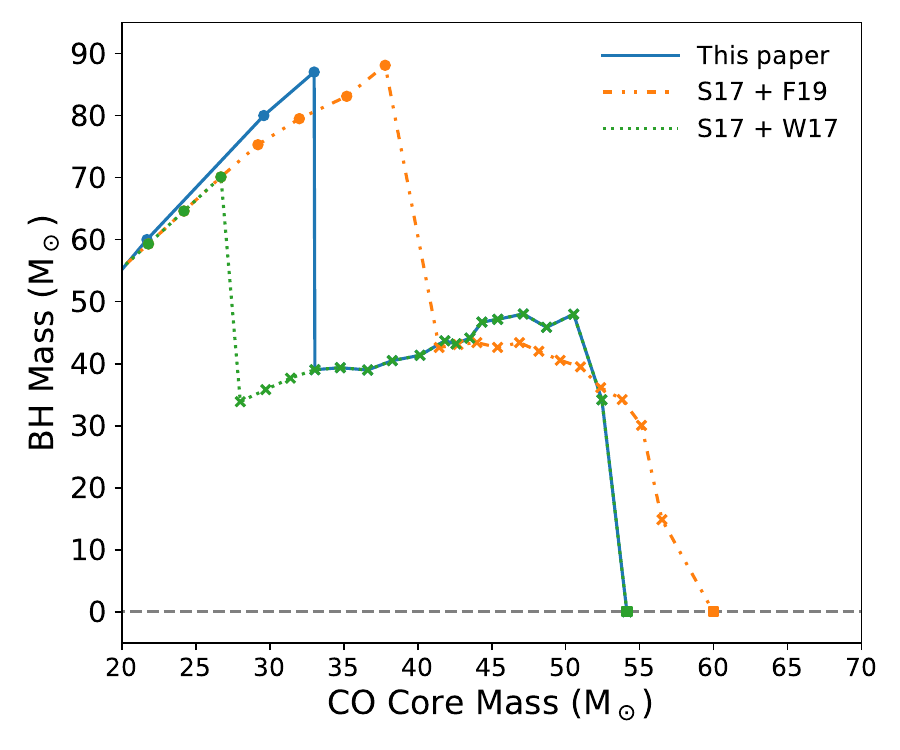}
    \caption{BH mass spectrum as a function of the CO core mass for different PPISN prescriptions. Solid blue line: this work, non-rotating progenitors at $\met =-3$; dash-dotted orange line: F19 PPISN threshold ($\rm M_{CO}^{PPISN} = 41\,\Ms$) applied to the S17 pre-SN progenitors; dotted green line: W17 PPISN threshold ($\rm M_{CO}^{PPISN} = 28\,\Ms$) applied to the S17 pre-SN progenitors. Circles (crosses) represent the models that
    are stable (unstable) against pair production, and squares represent the first star exploding as PISNe.}
    \label{fig: Varying PPI threshold}
\end{figure}

Figure \ref{fig: Varying PPI threshold} shows the BH mass spectrum of this work and the results we would obtain by applying the W17 and F19 criteria for the onset of PPISNe to the S17 stellar evolution tracks. It shows that even the progenitors from S17, which have the lowest PPISNe entry point ($\rm M_{CO}^{PPISN} = 24\,\Ms$), can form BHs as massive as $\sim 70\,\Ms$ adopting the W17 threshold for the CO core mass ($\rm M_{CO}^{PPISN} = 28\,\Ms$) instead of the He core mass criterion. Furthermore, S17 can form BH as massive as $\sim 90\,\Ms$ adopting the F19 threshold ($\rm M_{CO}^{PPISN} = 41\,\Ms$), without changing the C/O ratio at the He core depletion, which is one of the key parameters to determine the onset of PPISNe (as shown in F19).\par

Therefore, adopting a self-consistent criterion to enter the PPISNe regime is crucial to investigate the BH mass spectrum and to constrain the lower edge of the upper mass gap.
It is worth noting that the value of $\rm M_{ CO}^{PPI}$ is not the only property that affects the position of the lower edge of the upper mass gap. Besides the roles of DC, stellar winds, and rotation (cf. Sec. \ref{sec: Results}), the convective core overshooting parameter ($\alpha_{\rm ov}$) also has a central role in the formation of heavy stellar BHs. 
This is because overshooting dredges extra H from the envelope into the core during the central H burning, thus it governs the mass of the He core, and  - in turn - the mass of the CO core. 
\cite{Vink2021} showed that a small value of overshooting ($\alpha_{\rm ov} \lesssim 0.1$) is associated with smaller cores and stars likely end their life as Blue Super Giants (BSGs). Such stars are very compact, they tend to avoid the PPISNe regime, because they form smaller CO cores and they experience negligible mass loss during their life. As a consequence, such stars are likely to form very heavy BHs in the range $74\text{\textemdash}87\,\Ms$.
In contrast, stars with higher values of $\alpha_{\rm ov}$ exhibit more massive cores and likely evolve through the RSG phase, experiencing significant mass-loss episode. Hence, the stars enter the PPISNe with smaller pre-SN mass, forming less massive BHs.\par
We do not have conclusive evidences on the value of the core overshooting parameter yet, with observations providing evidence for higher possible values of the overshooting parameter than the one used in \cite{Vink2021} (e.g.  \citealt{Vink2010, Brott2011, Higgins2019, Bowman2020, Sabhahit2023, Winch2024}).
Furthermore, other authors such as \cite{Takahashi2018, Farmer2019, Farag2022, Mehta2022, Shen2023} show that the variation in the C/O ratio at the He core depletion plays a role in the onset of the PPISNe.\par
The parameter space of the factors ruling the final fate of massive stars is still mostly unexplored, and from this analysis it is apparent that there is still room for many future improvements. We will explore the impact of these different effects on the onset of PPISNe in a follow-up work.
\subsection{Remnants mass distribution}
\begin{figure*}[htb!]
    \centering
    \includegraphics[width=\textwidth]{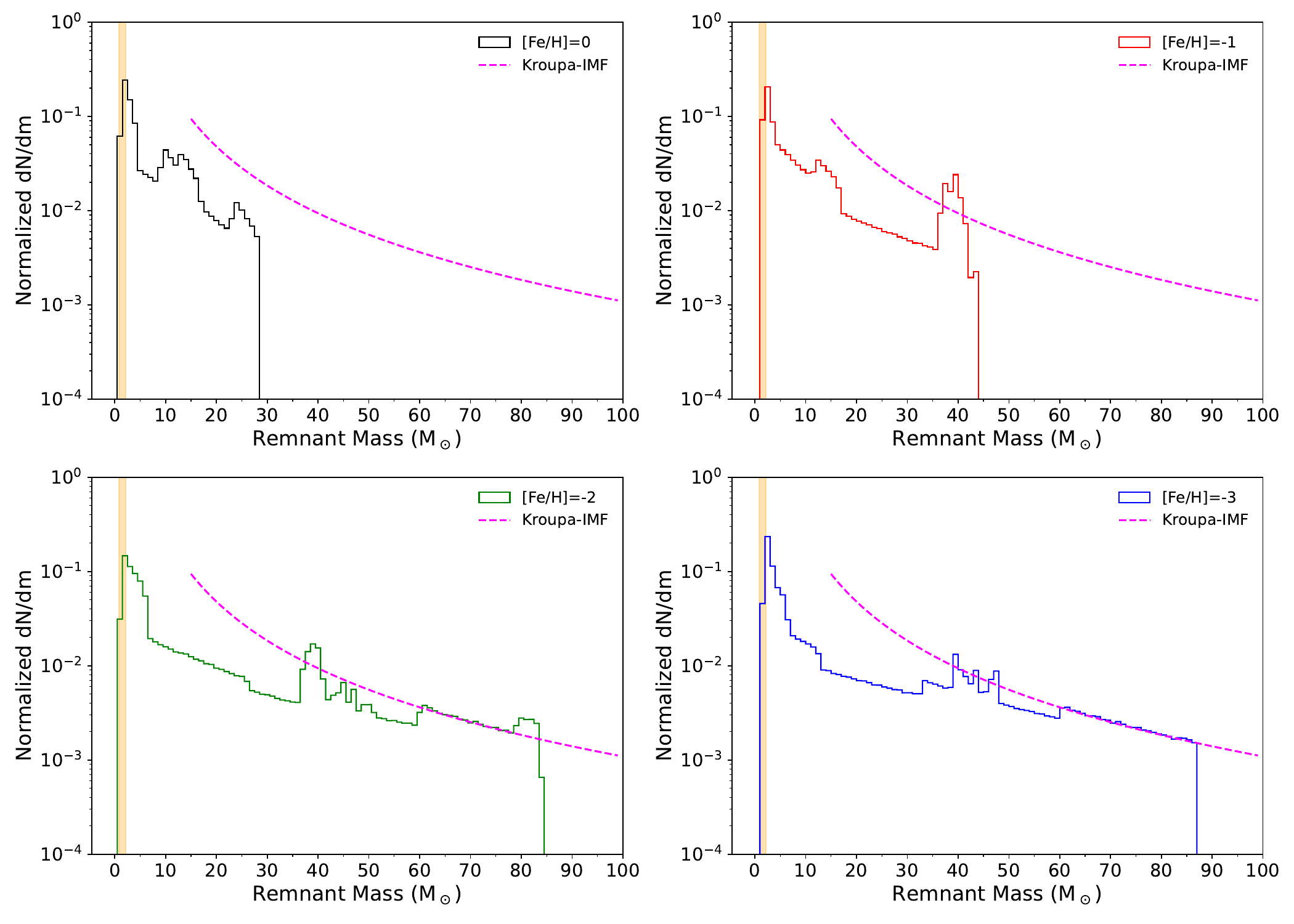}
    \caption{Normalized probability distribution of BH masses we obtain from a population of $10^6$ stars whose mass follows a \cite{Kroupa2001} mass function ($dN/ d\Mzam\propto \Mzam^{-2.35}$, dashed magenta line in all the panels) in the range $\Mzam \in \left[15\,\Ms; 150\,\Ms\right]$ for non-rotating stellar progenitors. The yellow shaded area represents the mass interval of NSs assuming $M_{\rm NS} \in \left[0.8\,\Ms;2.2\,\Ms\right]$. The Figure shows the remnant mass distribution we obtain for $\met=0$ (upper left panel), $\met=-1$ (upper right panel), $\met=-2$ (lower left panel), and $\met=-3$ (lower right panel).}
    \label{fig: BH NR distro}
\end{figure*}
\begin{figure*}[htb!]
    \centering
    \includegraphics[width=\textwidth] {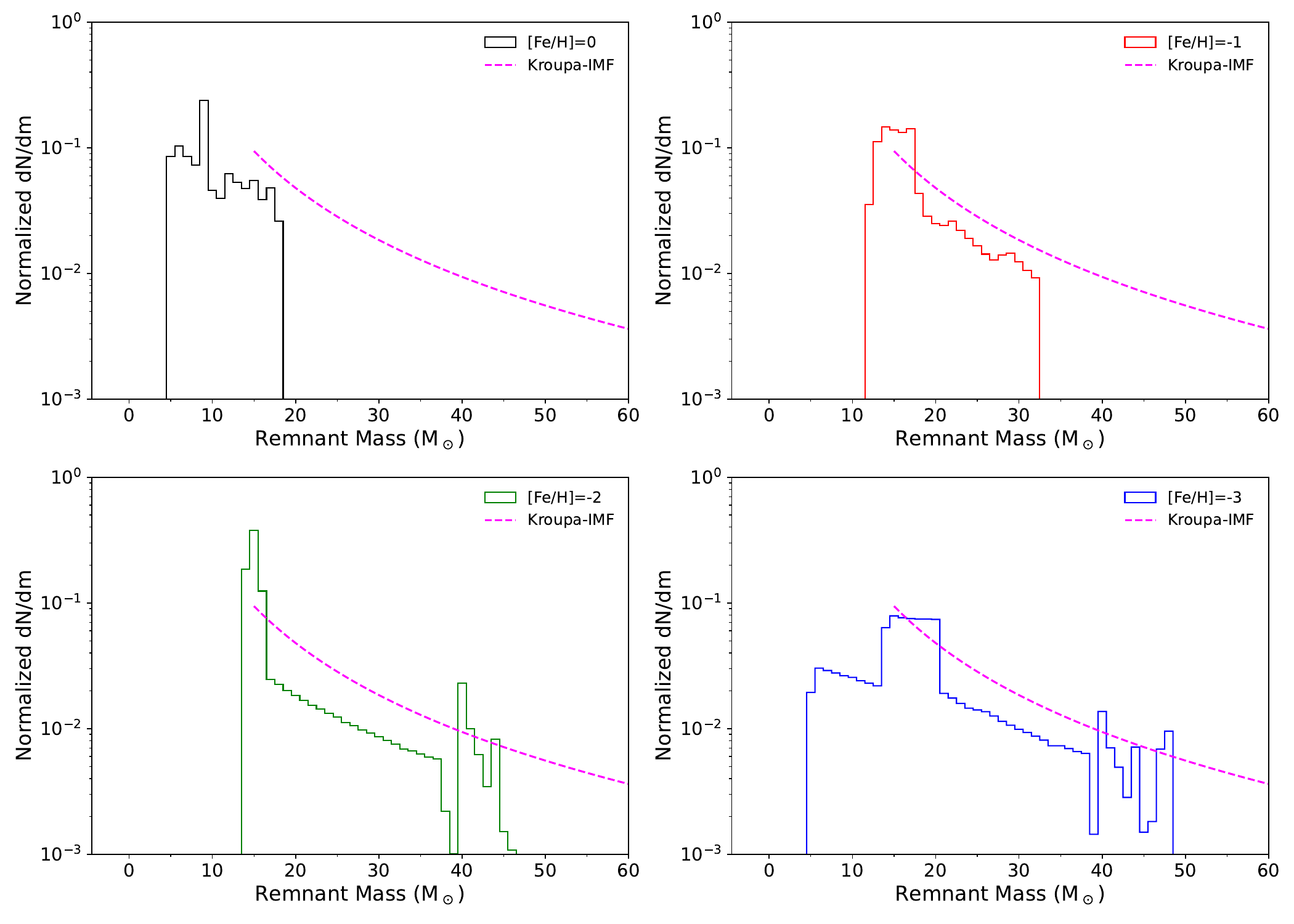}
    \caption{Same as Figure \ref{fig: BH NR distro} but for rotating progenitors.}
    \label{fig: BH R distro}
\end{figure*}

Figure \ref{fig: BH NR distro} and \ref{fig: BH R distro} show the remnants mass distribution we obtain by evolving a population of $10^6$ isolated single stars with an initial-mass function (IMF) $dN / d\Mzam \propto \Mzam^{-2.35}$ (\citealt{Kroupa2001}) and $\Mzam \in \left[15\,\Ms;150\,\Ms\right]$ for non-rotating and rotating progenitors, respectively.
The mass of the remnant formed for each values of $\met$ and $\Mzam$ is estimated by interpolating the $M_{\rm BH}\textrm{--}\Mzam$ curve we find in Sec. \ref{sec: non-rot progenitors}, \ref{sec: rot progenitors}.\par
The panels of Figure \ref{fig: BH NR distro} show several peaks in the remnants mass distribution. The most apparent peak of the remnants mass distributions is the one that corresponds to compact objects with mass $0.8\text{\textemdash}2\,\Ms$. This peak is a feature of the IMF and it shows up for all the considered metallicities for the non-rotating stars. It emerges because the majority of stellar progenitors extracted from the Kroupa IMF lie in the low-mass end with $\Mzam \in \left[15\,\Ms; 20\,\Ms\right]$, and they are expected to form NSs, as result of successful ccSNe.\par
At $\met=0$ (the top right panel of Figure \ref{fig: BH NR distro}), we find two other peaks in the remnant mass distribution, at BH masses of $\sim 9\text{\textemdash13}\,\Ms$ and $\sim 23\,\Ms$. The first one shows up because the $M_{\rm BH} \textrm{--} \Mzam$ curve flattens out for $\Mzam \in \left[30;40\right]$ causing an accumulation of BHs at $\sim 13\,\Ms$ (see the top left panel of Figure \ref{fig: Mpre-SN and Mrem vs Mini NR case}). The peak at $\sim 23\,\Ms$ occurs at the transition from successfully-exploding progenitors to those that directly collapse to BHs. 
The heaviest BHs we obtain have masses $\simeq 28 \, \Ms$, consistent with the results presented in Sec. \ref{sec: non-rot progenitors}.
At $\met=-1$ (top right panel of Figure \ref{fig: BH NR distro}) we find two peaks in the remnants mass distribution at $\sim 13\,\Ms$ and $\sim 37 -  43\,\Ms$. The former corresponds to a local flattening of the $M_{\rm BH} \textrm{--} \Mzam$ curve in the region of progenitors with $\Mzam=30 - 40\,\Ms$, and the latter is a degenerate effect due to the combination of the local flattening of the $M_{\rm BH} \textrm{--} \Mzam$ curve and to the  effect of the onset of the PPISNe, which causes significant mass loss and produce BHs with mass $\sim 37 - 48\,\Ms$. 
Furthermore, at this value of metallicity the heaviest BH that forms has a mass $\sim 44\,\Ms$, in agreement with the results presented in Sec. \ref{sec: PI regime}, \ref{sec: non-rot progenitors}.\par
At $\met=-2$ (bottom left panel of Figure \ref{fig: BH NR distro}), we find two peaks in the distribution for BHs with masses of $\sim 40\,\Ms$, which again corresponds to the piling-up of BHs we obtain from the PPISNe, and of $\sim 78 - 84\,\Ms $. This latter is the result of the piling-up of BHs from the stellar progenitors with $\Mzam$ between $80\,\Ms$ and $87\,\Ms$ where the slope of the $M_{\rm BH} \textrm{--} \Mzam$  curve decreases. 
Furthermore, the heaviest BHs that form have mass $\sim 83\,\Ms$.\par
As in the case $\met=-2$, at $\met=-3$ (bottom right panel of Figure \ref{fig: BH NR distro}) we find that BHs pile up at masses $\sim 40 - 43\,\Ms$ because of PPISNe. We find that above the threshold for DC, the BH mass distribution follows the IMF distribution, since stellar winds are quenched and stars directly collapse to BHs with mass equal to $\Mzam$.\par
In summary, the features emerging from the BH mass distribution significantly depend on the adopted model, including the prescriptions for the pulsational pair-instability, and  the details of the SN explosion. Furthermore, we find that the peak in the probability distribution corresponding to the transition between ccSNe and DC overlaps with the peak due to the PPISNe for some metallicities (e.g. $\met=-1$). Such peaks in the remnants mass distribution are not only uncertain and model dependent, but also degenerate with the details of the SN explosion mechanism.\par
From Figure \ref{fig: BH R distro} it is apparent that stellar rotation significantly affects the remnants mass distribution. The peak corresponding to stellar progenitors extracted from the low-mass end of the Kroupa IMF is now shifted to BH with masses $\sim 4\,\Ms$ for $\met =0,\,-3$ (upper left panel and bottom right panel of Figure \ref{fig: BH R distro}, respectively) and $\sim 12\,\Ms$  for $\met=-1,\,-2$ (upper right panel and bottom left panel of Figure \ref{fig: BH R distro}, respectively). This effect is a degenerate feature of the abundance of stars with $\Mzam =15 \text{---} 20\,\Ms$ and the outcome of the SN explosions at different initial metallicities for these stellar progenitors (see Section \ref{sec: rot progenitors}). 
Thus, we find that rotating stellar progenitors cannot produce NSs. 
At $\met=0$, we find that the distribution has a peak corresponding to $\sim 8\,\Ms$, which is due to the local flattening of the $M_{\rm BH} \textrm{--} \Mzam$ curve in the interval of $\Mzam$ between $20\,\Ms$ and $25\,\Ms$. The heaviest BH we form at solar metallicity has mass $\sim 19\,\Ms$.\par

At $\met=-2$ we find a peak in the remnants mass distribution corresponding to BHs masses of $\sim 40\,\Ms$. This peak comes from the onset of the PPISNe. Such feature is independent of the details of the SN explosion and depends only on the prescription for the results of the PPISNe (combination of LC18 and W17 in this work).
At $\met=-3$, we find a peak in the remnants mass distribution, which corresponds to BH masses of $\sim 13 - 20\,\Ms$. The piling-up of BHs in this mass region is due to the particular shape of the $M_{\rm BH} \textrm{--} \Mzam$ curve in the region of $\Mzam$ between $15\,\Ms$ and $\sim 30 - 40\,\Ms$. In such region, the $M_{\rm BH} \textrm{--} \Mzam$ curve has a local minimum at $\Mzam=25\,\Ms$, which produce a BH with mass $\sim 13\,\Ms$. Thus, a wide interval of progenitors piles-up BHs with mass in the interval $\sim 13 - 20\,\Ms$ (see also Figure \ref{fig: Mpre-SN and Mrem vs Mini R case}). Furthermore, we find peaks in the remnant mass distribution corresponding to BH masses of $\sim 40\,\Ms$ $\sim 43\,\Ms$ and $\sim 47\,\Ms$. The first peak is a degenerate feature from both the SN explosion details and the prescriptions for the PPISNe. The latter two depend only on the $\Mzam\text{---}M_{rem}$ curve we assume for the PPISNe.
Finally, it is worth stressing that our results are based only on isolated single stars. Thus, considering binary stellar systems may significantly affects the results one can found about the BH mass spectrum.

\subsection{Prescriptions for Population Synthesis Codes}
\begin{figure*}[htb]
    \centering
    \includegraphics[width=\columnwidth]{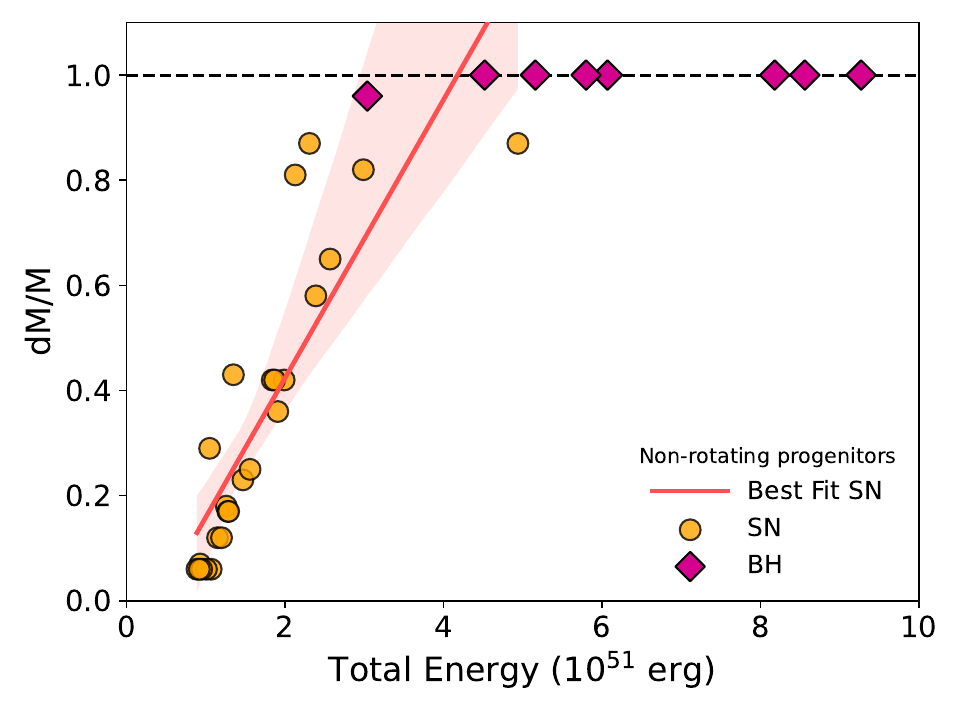}\quad
    \includegraphics[width=\columnwidth]{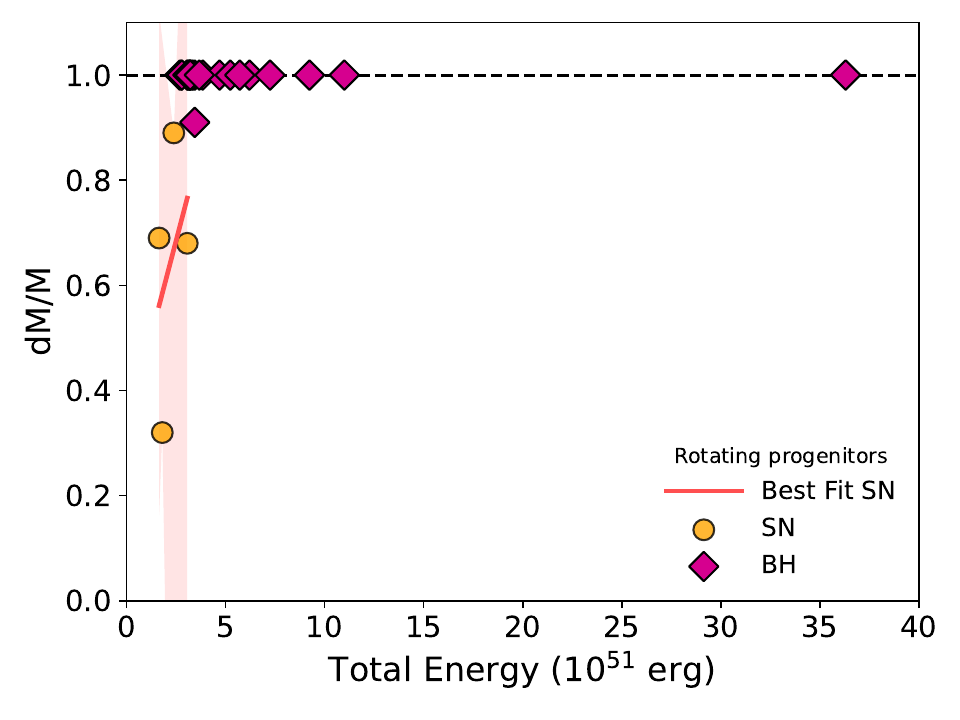}\\
    \includegraphics[width=\columnwidth]{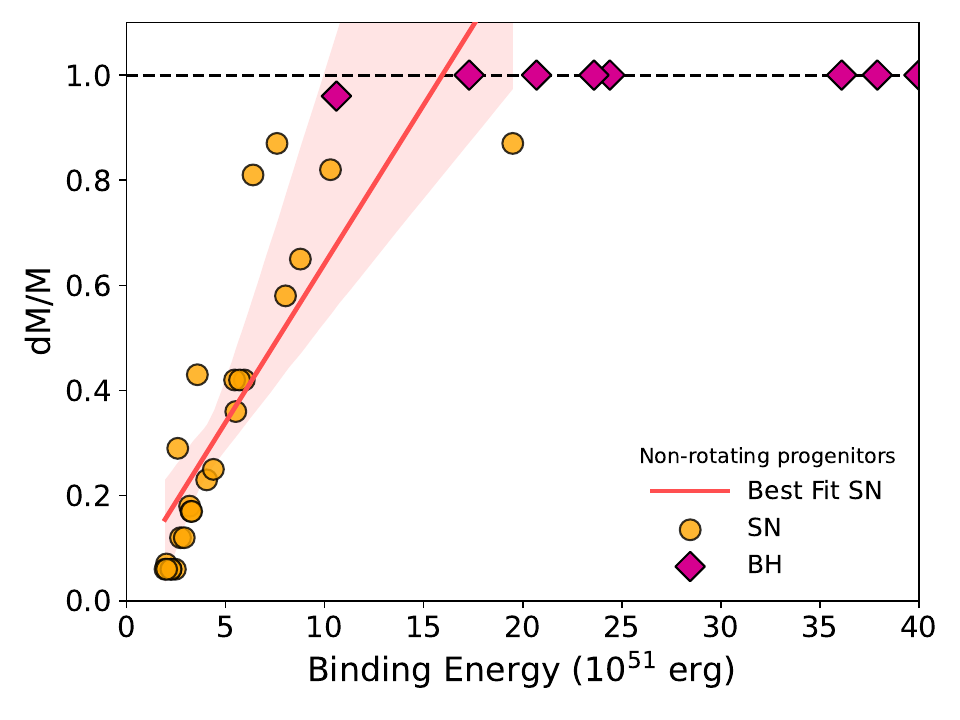}\quad
    \includegraphics[width=\columnwidth]{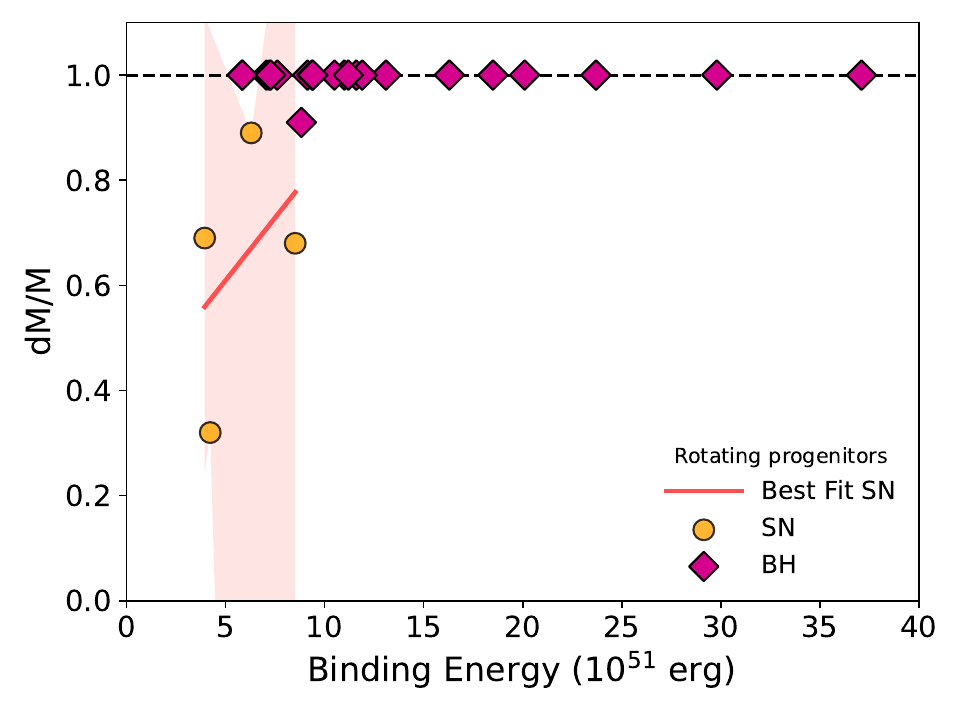}\\
    \includegraphics[width=\columnwidth]{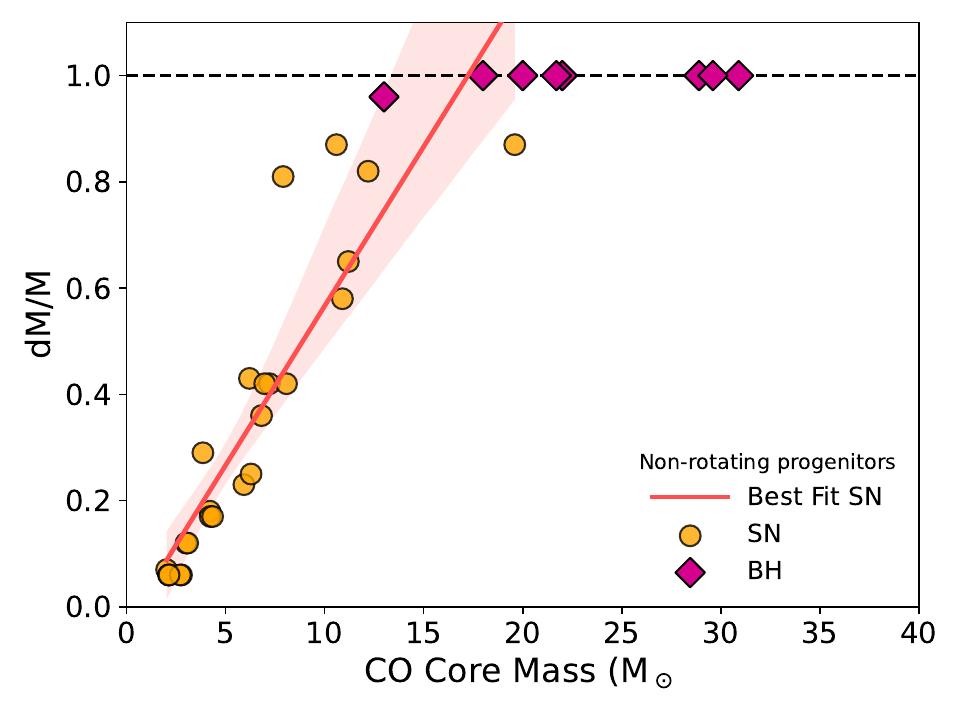}\quad
    \includegraphics[width=\columnwidth]{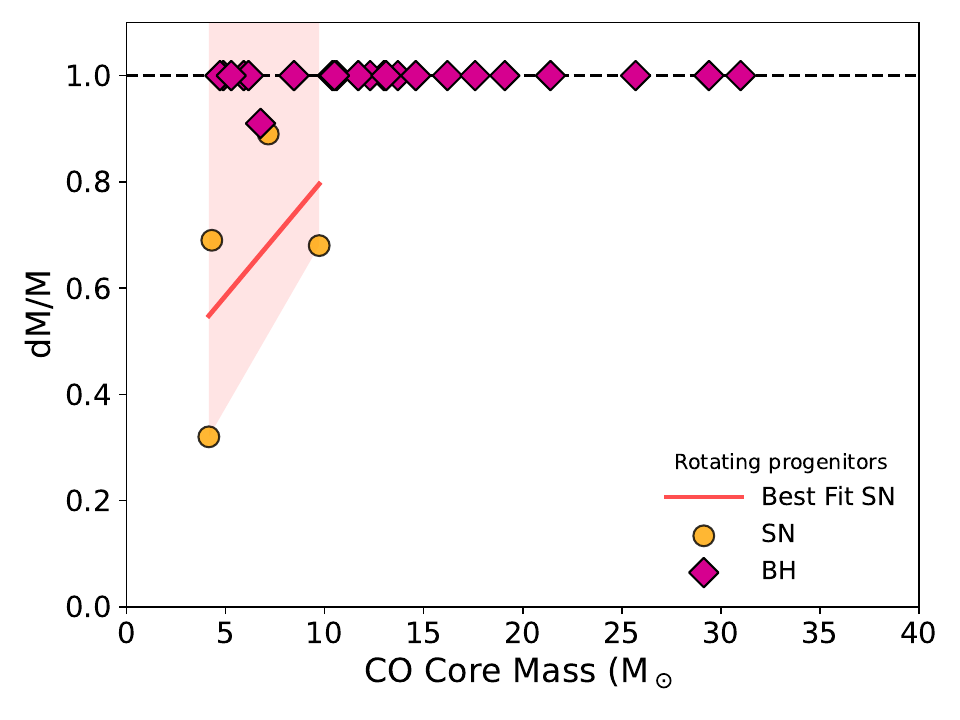}\\
    \caption{Linear relation between the mass that falls back after the explosions and some structural properties of the stellar progenitors at the pre-SN stage, such as the absolute value of the total energy (top panels), the binding energy (central panels) and the CO core mass (bottom panels). We have performed such analysis both for the non-rotating (left panels) and for the rotating stellar progenitors (right panels). We study the fate of the explosion independently of the initial metallicity, thus in each panel we have all the models with $\met=0\,-1,\,-2,\,-3$. In each panel the circular orange points are the stellar progenitors undergoing successful SN explosions and the best fit for $dM/M$ is the solid red line, while the diamond purple points represents the progenitors which DC to BHs. The black dashed line corresponds to $dM/M=1$ i.e., the value of dM/M for the DC. The red shaded area is the 95\% confidency interval for the SN linear regression.}
    \label{fig: fit for DM/M}
\end{figure*}

In this section, we present a set of analytic prescriptions derived from our results. Such fitting formulae can be implemented in population-synthesis models to calculate the mass of the remnants, given the physical properties of the progenitor at the pre-SN stage.\par

The main value that our prescriptions provide is the fraction of mass that falls back onto the compact remnant during the explosion $dM/M$, that is:
\begin{equation}
    \dfrac{dM}{M}= \dfrac{M_{\rm rem}}{M_{\rm preSN}}
\end{equation}
with $M_{\rm rem}$ that is the final remnant mass we obtain from the explosion.\par
The fraction of ejected mass is parametrized as a function of either the absolute value of the total energy, the binding energy of the stellar progenitor, or the CO core mass at the pre-SN stage. The total energy is computed as:
\begin{equation}
    E_{tot}=E_{bind}+E_{int}+E_{kin}
\end{equation}
where $E_{bind}$ is the binding energy of the whole progenitor, $E_{int}$ is the internal energy and $E_{kin}$ is the kinetic energy of the stellar matter, over the whole progenitor.

The panels of Figure \ref{fig: fit for DM/M} show the linear fit we obtain for the non-rotating (left panels) and the rotating progenitors (right panels), as a function of the absolute value of the total energy (top), binding energy (middle), and CO core mass (bottom). For such analysis we study the fate of the explosion independently of the initial metallicity, thus in each panel we have all the models with $\met=0\,-1,\,-2,\,-3$.\par

We first consider non rotating progenitor models. We find that we can fit the relation between $dM/M$ and the absolute value of the total energy as:

\begin{align}
    dM/M=
   \begin{cases}
    a_{\rm SN}\left|E_{\rm tot}\right|+b_{\rm SN},  & \left|E_{\rm tot}\right| < \left|E_{\rm tot}\right|_{\rm DC} \\
    0, &\left|E_{\rm tot}\right| \geq \left|E_{\rm tot}\right|_{\rm DC}
    \end{cases}
\end{align}  
with $a_{\rm SN}= 0.26\, \rm foe^{-1}$, and $b_{\rm SN}=-0.11$, and $\left|E_{\rm tot}\right|_{\rm DC}=4.17\, \rm foe$. The latter is the value of $\left|E_{tot}\right|$ for which $dM/M=0$, i.e. for the DC.\par
From the top-left panel of Figure \ref{fig: fit for DM/M} it is apparent that two progenitors deviate significantly from our best fit. We find one stellar progenitor that directly collapses to BH despite having  $\left|E_{\rm tot}\right|<\left|E_{\rm tot}\right|_{\rm DC}$, and one progenitor that explodes with $\left|E_{\rm tot}\right|>\left|E_{\rm tot}\right|_{\rm DC}$. The former star has $\Mzam=60\,\Ms$ and $\met=0$ and ends its life as a He-star, while the latter has $\Mzam=60\,\Ms$ and $\met=-1$, and ends its life as a RSG, and therefore its H envelope is very loosely bound despite the high absolute value of the total energy. 
The relation between $dM/M$ and the binding energy $E_{\rm bind}$ can be fit as:
\begin{align}
    dM/M=
   \begin{cases}
    c_{\rm SN}E_{\rm bind}+d_{\rm SN},  & E_{\rm bind} < E_{\rm bind,\rm DC} \\
    0, &E_{\rm bind} \geq E_{\rm bind, DC}
    \end{cases}
\end{align} 
With $c_{\rm SN}= 0.06\, \rm foe^{-1}$, and $d_{\rm SN}=0.04$, and $E_{\rm bind, DC}=15.9\, \rm foe$.\par
From the central-left panel of Figure \ref{fig: fit for DM/M} it is apparent that one progenitor deviates significantly from the best fit. This is again the model with $\met=-1$ and $\Mzam=60\,\Ms$, which explode as a SN with a binding energy of $E_{bind}=19.5\,\rm foe$.
If we did not consider this stellar progenitor, we would find a threshold value of $\sim 10.0 \, \rm foe$ for the binding energy above which the stars DC to BHs.\par

Finally, the relation between $dM/M$ and the CO core mass $M_{\rm CO}$ can be fit as: 
\begin{align}
 dM/M=
   \begin{cases}
    e_{\rm SN}M_{\rm CO}+f_{\rm SN},  & M_{\rm CO} <M_{\rm CO, DC} \\
    0, &M_{\rm CO} \geq M_{\rm CO, DC}
    \end{cases}   
\end{align} 

With $e_{\rm SN}= 0.06\, \Ms^{-1}$, and $\boldsymbol{f_{\rm SN}=-0.03}$, and $M_{\rm CO}^{\rm DC}=17.2\, \Ms$, where again we find that only the two progenitor stars with $\Mzam=60\,\Ms$, $\met=-1$ (that explodes as a SN despite having $M_{\rm CO} = 21.4\,\Ms > M_{\rm CO, DC}$), and with $\Mzam=60\,\Ms$, $\met=0$ (the He-star with $M_{\rm CO} = 16.6\,\Ms < M_{\rm CO, DC}$ that directly collapses into a BH) deviate from the global trend.\par

In contrast, the right panels of Figure \ref{fig: fit for DM/M} show that rotating progenitors do not follow a linear relation between $dM/M$ and $\left|E_{\rm tot}\right|$, $E_{\rm bind}$, or  $M_{\rm CO}$, and that there are no threshold values of any of these quantities that clearly separate stars which explode as SNe from stars which DC to BHs. Therefore, we cannot find any linear relation between the properties of rotating stars at the pre-SN stage and the final-remnant mass.\par

It is important to stress that - even in the case of non-rotating stellar progenitors - we can only provide a rough estimate of the thresholds for the transition from stars that successfully explode as ccSNe and stars that directly collapse to BH. From the left panels of Figure \ref{fig: fit for DM/M}, it is apparent that a linear relation is in place below the threshold for the DC, but that some stars deviate from the expected trends, indicating that this approach fails to fully capture the complexity of a SN explosion. Our analysis shows that the fate of the collapse of each  stellar progenitor is ruled only by the dynamics of the shock wave, i.e. the  micro-physics effects taking place during the explosion are crucial in determining its outcome.\par

\section{Summary and Conclusion}\label{sec: summary and conclusions}

In this paper, we studied the explosion of a subset of pre-SN models of massive stars presented in \cite{limongi_presupernova_2018}. The subset contains massive stars with masses in the range $13 -120\,\Ms$, initial metallicities $\met=0,\,-1,\,-2,\,-3 $, and initial rotation velocities $v=0\, \rm km\, s^{-1},\,300\, \rm km\, s^{-1}$. Using the HYPERION code \citep[][]{limongi_hydrodynamical_2020}, we simulate the explosion by removing the inner 0.8 $\Ms$ of the iron core, and injecting a given amount of thermal energy at the edge of this zone. Such approach required a set of free parameters to model the injection of energy. To fix these parameters, we have performed several calibration runs of a stellar progenitor similar to the progenitor of SN 1987A (Sk — 69°202), aimed at reproducing the observed kinetic energy of the ejecta ($\sim 1.0 \, \rm foe$) and $\nickel$ mass ($\sim 0.07 \, \Ms$) \citep{Arnett1989, Utrobin1993, Utrobin2005, Utrobin2006, Blinnikov2000, Shigeyama1990}.

Once we have chosen the explosion parameters, we have adopted this combination for the explosion of all the models of our grid of stellar progenitor, obtaining both successful and failed explosions (DC to BHs), depending on the initial stellar parameters, i.e. mass, metallicity and rotation velocity.\par

We find that:
\begin{itemize}
\item The metallicity plays a crucial role in the pre-SN evolution of a star, since it is strongly connected with the efficiency of stellar winds. For non rotating models, progenitors with $\met=-2$ and $\met=-3$ evolve at roughly constant mass during their life, while for $\met=0$ and $\met=-1$ the stellar winds remove a significant fraction of mass from the stellar progenitor. For these models we find that the threshold masses above which the stars undergo a DC to BH are  $\Mzam \simeq 60\,\Ms$ for $\met=0,\, -2,\, -3$ and $\Mzam \simeq 80\,\Ms$ for $\met=-1$. The most massive BH has mass of $\simeq 91 \,\Ms$, and comes from the DC of the most massive star at $\met=-3$ not undergoing PPISNe (assuming that $\rm M_{\rm CO}^{PPISN} = 33 M_\odot$, see below).

\item The presence of an initial angular rotation enhances the mass loss, and induces the formation of larger convective regions, and therefore larger CO cores, i.e. more bound structures. 
Thus, we find that for rotating stellar progenitors, the transition from a successful SN explosion to a DC to BHs occurs at lower $\Mzam$ values ($\sim 20\,\Ms$, $\sim 30\,\Ms$, $\sim 15\,\Ms$, and $\sim 20 \,\Ms$, for $\met=0$, $\met=-1$, $\met=-2$, and $\met=-3$, respectively) than for the non-rotating progenitors.  In contrast, because of the enhanced mass loss driven by rotation, even if more stars DC to BHs, the maximum BH masses formed are lower than for non-rotating stars. In addition, the chemical mixing induces the onset of the PPISNe at lower values of $\Mzam$ with respect to the non-rotating progenitors. As a result, the maximum BH mass formed by rotating progenitors is $\sim 41\,\Ms$, which comes from the DC of a stellar progenitors undergoing PPISNe, that we model according to the results of W17 (assuming that $ \mathrm{M_{\rm CO}^{PPISN}} = 33\,\Ms$, see below).

\item The maximum BH mass formed is very sensitive to the adopted criterium for the onset of PPISNe, which can be expressed in terms of the maximum CO core mass above which stars enter the PPISN regime, $\rm M_{\rm CO}^{PPISN}$. From our simulated grid we find that $30 M_\odot < M_{\rm CO}^{\rm PPISN} < 50 M_\odot$, which is consistent with most of the published results. However, these vary considerably, from the lowest value adopted by S17, $\rm M_{CO}^{PPISN} = 24\,\Ms$, to the highest value adopted by F19, $M_{\rm CO}^{\rm PPISN}=41\,\Ms$. 
Such a discrepancy results in a significant uncertainty on the maximum BH mass, which can be as low as $\sim 53\,\Ms $ in the case of S17, or even as high as $\sim 87\,\Ms$, which is the case of this paper.

\item We have applied the results of our simulations to a populations of $10^6$ stars whose masses are distributed accordingly to a Kroupa-IMF. The resulting BH probability distribution depends on the initial stellar metallicity and rotation rate, and encode information on the progenitor mass - BH mass relation. We showed that the features emerging in the BH mass distribution depend significantly on the adopted stellar models and show significant degeneracies.

\item We provide linear relationships between the mass ejected during the explosion $dM/M$ and the structural properties of the stellar progenitors at the pre-SN stage, such as the binding energy, the CO core mass and the absolute value of the total energy. Such relationships have been found only in the case of non-rotating stellar progenitors, even if some stars deviates from the linear fit.
In contrast, we do not find any monotonic relationship between $dM/M$ and the progenitor's properties in the case of rotating stellar progenitors. We showed that, despite the relations can be adopted for fast population-synthesis studies, they fail to capture the details of the link between progenitor stars and their compact remnants.

\end{itemize}

It is worth mentioning that in this work we calibrate the explosions on the observational properties of SN1987A (i.e. the kinetic energy of the ejecta, and the mass of $\nickel$ ejected into the interstellar medium). In a follow-up study we will broaden the set of SNe used for the calibration process.\par
Furthermore, we plan to increase the resolution in our grid of stellar progenitors in terms of $\Mzam$, initial metallicity and initial angular rotation to produce self-consistent estimates of the remnant masses of a larger grid of pre-SN models.\par
Finally, we plan to develop a large set of look-up tables, including rotating progenitor stars, that can be implemented in rapid binary population synthesis codes (e.g. SEVN \citealt{Spera2015, Spera2019, Spera_2017}) to investigate the evolution of compact-object binaries and the implications on detectable gravitational-wave signals.


\begin{acknowledgements}
This work has been partially developed at The Unconventional Thinking Tank Conference 2022, which is supported by INAF.
MS acknowledges financial support from Fondazione ICSC, Spoke 3 Astrophysics and Cosmos Observations. National Recovery and Resilience Plan (Piano Nazionale di Ripresa e Resilienza, PNRR) Project ID CN\_00000013 ``Italian Research Center on High-Performance Computing, Big Data and Quantum Computing'' funded by MUR Missione 4 Componente 2 Investimento 1.4: Potenziamento strutture di ricerca e creazione di ``campioni nazionali di R\&S (M4C2-19 )'' - Next Generation EU (NGEU), and from the program ``Data Science methods for Multi-Messenger Astrophysics \& Multi-Survey Cosmology'' funded by the Italian Ministry of University and Research, Programmazione triennale 2021/2023 (DM n.2503 dd. 09/12/2019), Programma Congiunto Scuole.
RS acknowledges support from the PRIN 2022 MUR project 2022CB3PJ3 - First Light And Galaxy aSsembly (FLAGS) funded by the European Union – Next Generation EU and from ICSC – Centro Nazionale di Ricerca in High Performance Computing, Big Data and Quantum Computing,
funded by European Union – NextGenerationEU.
\end{acknowledgements}

\bibliographystyle{aa} 
\bibliography{References}

\appendix

\section{The FRANEC code}

In this Appendix, we summarize the main features of the FRANEC stellar evolutionary code. For a complete description, we refer the reader to \cite{chieffi_pre-supernova_2013} and \cite{limongi_presupernova_2018}.\par
The FRANEC code couples all the equations describing the physical structure of the star and the chemical evolution of the matted due to both the nuclear ractions and the various mixing processes (convection, rotation, diffusion, semiconvectionm etc.) and solves them simultaneously by using a relaxation technique.\par
Convective envelope boundaries are determined using the Ledoux criterion, and an overshooting of $0.2 H_p$ is included at the convective core's outer edge during the core hydrogen-burning phase.\par
The equation of state (EoS) provided by \cite{Rogers1996, Rogers2001} (EOS and EOSPLUS) is adopted for temperatures below $10^6 \, \mathrm{K}$ while the EoS provided by \cite{Straniero1988} is used for temperatures $T \geq 10^6 \, \mathrm{K}$. Radiative opacity coefficients are derived from \cite{Kurucz1991} for $T \leq 10^4 \, \mathrm{K}$, \cite{Iglesias1992} (OPAL) and the Los Alamos Opacity Library (LAOL) \cite{Huebner_1977} for $10^4 \, \mathrm{K} < T \leq 10^8 \, \mathrm{K}$, and $T \geq 10^8 \, \mathrm{K}$, respectively. Thermal conductivity opacity coefficients follow \cite{Itoh1983}.\par
Stellar wind mass loss is implemented according to \cite{vink_new_2000, Vink2001} for the blue supergiant (BSG) phase ($T_\mathrm{eff} \geq 1.2 \times 10^4 \, \mathrm{K}$), \cite{deJager1988} for the red supergiant (RSG) phase ($T_\mathrm{eff} < 1.2 \times 10^4 \, \mathrm{K}$), and \cite{Nugis2000} for the Wolf-Rayet (WR) phase. For stars in the RSG phase, dust-enhanced stellar winds are included following \cite{vanLoon2005}. In rotating models, mass loss is enhanced according to \cite{Heger2000}.\par
The classification criteria for WR subclasses follow \cite{Limongi2006}:
\begin{itemize}
    \item WNL: $10^{-5} < H_\mathrm{surf} < 0.4$,
    \item WNE: $H_\mathrm{surf} < 10^{-5}$ and $(C/N)_\mathrm{surf} < 0.1$,
    \item WNC: $0.1 < (C/N)_\mathrm{surf} < 10$,
    \item WCO: $(C/N)_\mathrm{surf} > 10$.
\end{itemize}

The nuclear network used in LC18 includes 335 isotopes, ranging from neutrons to $\mathrm{Bi}^{209}$, as detailed in Table 1 of LC18. For isotopes near the magic numbers $N = 82$ and $N = 126$, only neutron captures and beta decays are explicitly followed, while intermediate isotopes are assumed to be in local equilibrium. This approach ensures computational efficiency without compromising accuracy, as validated by one-zone core helium burning calculations comparing the full network to an extended one, both yielding consistent abundances around $A = 131-145$.\par
Nuclear cross-sections and weak interaction rates in LC18 are updated with respect to \cite{chieffi_pre-supernova_2013}, when possible, using the latest data. Most of the nuclear rates are extracted from the STARLIB database \cite{Sallaska2013}, with Tables 3 and 4 of LC18 providing a full reference matrix of all the processes included in the network along with their proper legend.\par
The effect of rotation on stellar structure is modeled using the "shellular rotation" approach (\citealt{Zahn1992, Meynet1997}). Angular momentum transport due to meridional circulation and shear turbulence is computed via an advective-diffusive equation \cite{Chaboyer1992, Talon1997}:
\begin{equation}\label{advdiffeq}
    \rho r^2 \frac{\partial (r^2 \omega)}{\partial t} = \frac{1}{5} \frac{\partial}{\partial r} \left( \rho r^4 U \omega \right) + \frac{\partial}{\partial r} \left( \rho r^4 D_\mathrm{s.i.} \frac{\partial \omega}{\partial r} \right),
\end{equation}
where $U$ is the radial velocity from meridional circulation, $D_\mathrm{s.i.}$ is the shear turbulence diffusion coefficient, $\omega$ is angular rotation, $\rho$ is density, and $r$ is radius. Prescriptions for $U$ are from \cite{Maeder1998}, and $D_\mathrm{s.i.}$ follows \cite{Palacios2003}:
\begin{equation}
    D_\mathrm{s.i.} = \frac{8}{5} \frac{R_\mathrm{ic} (r \, d\omega/dr)^2}{N_T^2/(K + D_h) + N_\mu^2/D_h},
\end{equation}
where \(N_T^2 = \frac{g}{H_p} (\nabla_{\text{ad}} - \nabla_{\text{rad}})\), \(N_\mu^2 = \frac{g}{H_p} \left( \frac{\varphi}{\delta} \nabla_\mu \right)\), \( R_{\text{ic}} = \frac{1}{4}\). $D_h = |r U|$ is horizontal turbulence (\citealt{Zahn1992}), and $K$ is thermal diffusivity (\citealt{Meynet2000}).\par
The adoption of the meridional circulation velocity provided by \cite{Maeder1998} transforms equation \ref{advdiffeq} into a fourth-order partial differential equation, that FRANEC solves with a relaxation technique.\par
Rotation driven mixing is treated as a purely diffusive process (\citealt{Chaboyer1992}). The diffusion coefficient is given by:
\begin{equation}
    D = f_c (D_\mathrm{s.i.} + D_\mathrm{m.c.}),
\end{equation}
where $D_\mathrm{m.c.} = |r U|^2/(30 D_h)$ (\citealt{Zahn1992}). Calibration parameters $f_c$ and $f_\mu$ adjust mixing and weight the molecular weight gradient. These calibrations are detailed in LC18.\par
Finally, the initial helium abundance and mixing-length parameters are calibrated to match the Sun's present-day properties.

\section{Calibration parameters}
\label{app: calibration parameters}

In Sec.\ref{sec: Calibration of the explosion} we present the calibration of our parameters and the final set we choose to stimulate the SN explosion in our simulations. In this Appendix we will discuss further the contribution of the different parameters $E_{\rm inj}$, $dm_{\rm inj}$ and $dt_{\rm inj}$.

\subsection{Calibration of $E_{inj}$}

\begin{figure}[htb]
    \centering
    \includegraphics[width=\columnwidth]{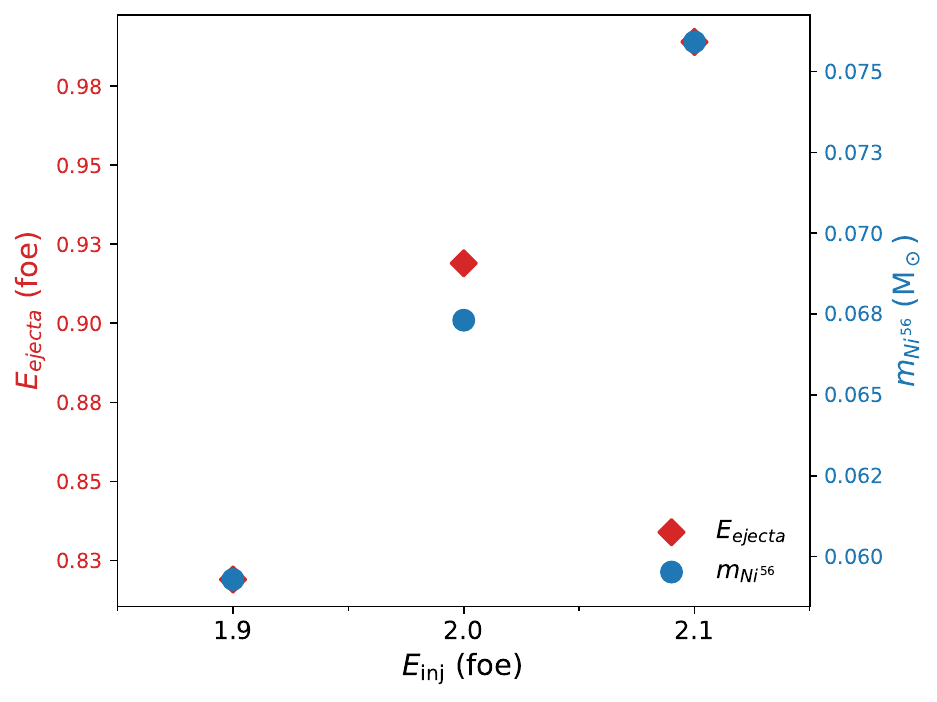}
    \caption{The Figure shows the values of $\Eejecta$ (red diamonds, whose scale is on the left-hand y-axes) and $\mnickel$ (blue points, whose scale is on the right-hand y-axis) as function of $E_{\rm inj}$.}
    \label{fig_app: Calibration E_inj}
\end{figure}

Figure \ref{fig_app: Calibration E_inj} shows the values that we find for $\Eejecta$ and $\mnickel$ for different values of $E_{\rm inj}$. We studied the variation of the outcomes of the explosion assuming $dt_{\rm inj}=10^{-9}s$ and $dm_{\rm inj}=0.1\Ms$, and by injecting the thermal energy at the arbitrary mass coordinate of $0.8\Ms$. The values we find for $\Eejecta$ and $\mnickel$ are also reported in Table \ref{tab_app: Calibration E_inj}.\par
We find that increasing the value of thermal energy inducing the explosion, $\Eejecta$ and $\mnickel$ increase as well. This is because the SN is stronger, thus the explosive nucleosynthesis reaches outer layers of the star, i.e. producing more $\nickel$.\par
We chose $E_{\rm inj}=2.0\, \rm foe$ because above this values we find an overproduction of $\nickel$, while the energy of the explosion is compatible with the value usually assumed in literature (see \citealt{Arnett1989, Shigeyama1990, Utrobin1993, Utrobin2006, Blinnikov2000}).

\subsection{Calibration of $dm_{\rm inj}$}

\begin{figure}[htb]
    \centering
    \includegraphics[width=\columnwidth]{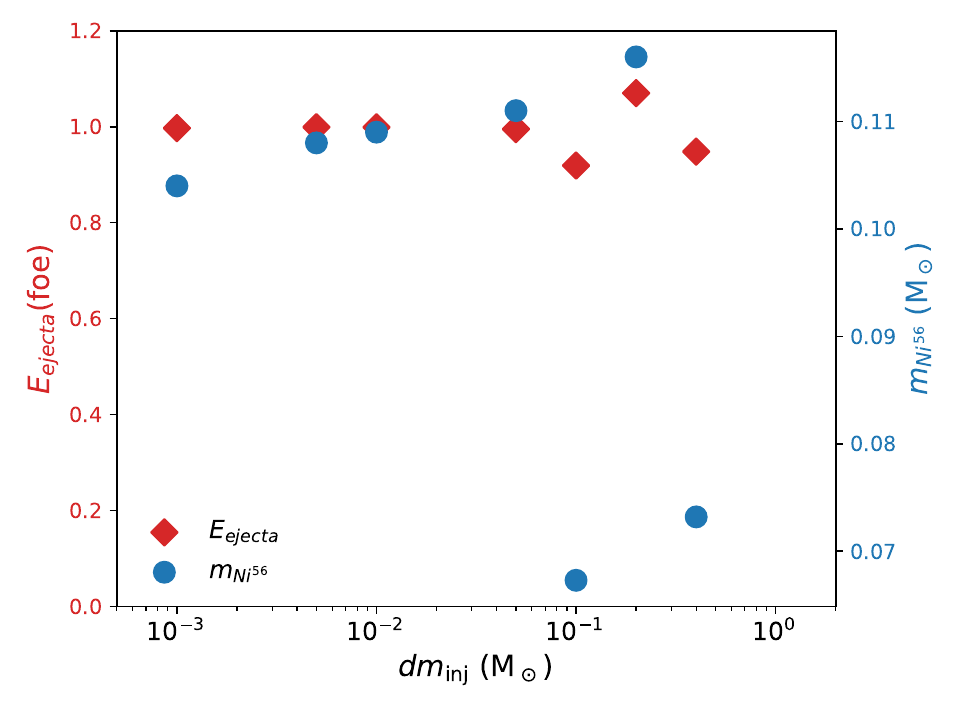}
    \caption{The Figure shows the values of $\Eejecta$ (red diamonds, whose scale is on the left-hand y-axes) and $\mnickel$ (blue points, whose scale is on the right-hand y-axis) as function of $dm_{\rm inj}$.}
    \label{fig_app: Calibration dm_inj}
\end{figure}

Figure \ref{fig_app: Calibration dm_inj} shows the values that we find for $\Eejecta$ and $\mnickel$ for different values of $dm_{\rm inj}$. We studied the variation of the outcomes of the explosion assuming $dt_{\rm inj}=10^{-9}s$ and $dE_{\rm inj}=2.0 \rm foe$, and by injecting the thermal energy at the arbitrary mass coordinate of $0.8\Ms$. The values we find for $\Eejecta$ and $\mnickel$ are also reported in Table \ref{tab_app: Calibration dm_inj}.\par
We find that if we choose to inject the energy in thinner layers, the explosion is more energetic, and because of this when the shock reaches the silicon shell, the explosive Si-burning is more efficient, with respect to scenarios with larger $dm_{\rm inj}$. Thus, in these cases we have an overproduction of $\nickel$ and the outcome of the explosion do not match the observations of SN1987A.\par
Our simulations are in agreement with the observations for $dm_{\rm inj}=0.1\Ms$ and $dm_{\rm inj}=0.4\Ms$. We chose $dm_{\rm inj}=0.1\Ms$, since our goal was to obtain a parameter set in good agreement with the observation of SN1987A, not the best parameter set because such analysis would have required a more sophisticated approach, and to calibrate on multiple SN sources. We will explore this aspect in an upcoming work.

\subsection{Calibration of $dt_{\rm inj}$}

\begin{figure}[htb]
    \centering
    \includegraphics[width=\columnwidth]{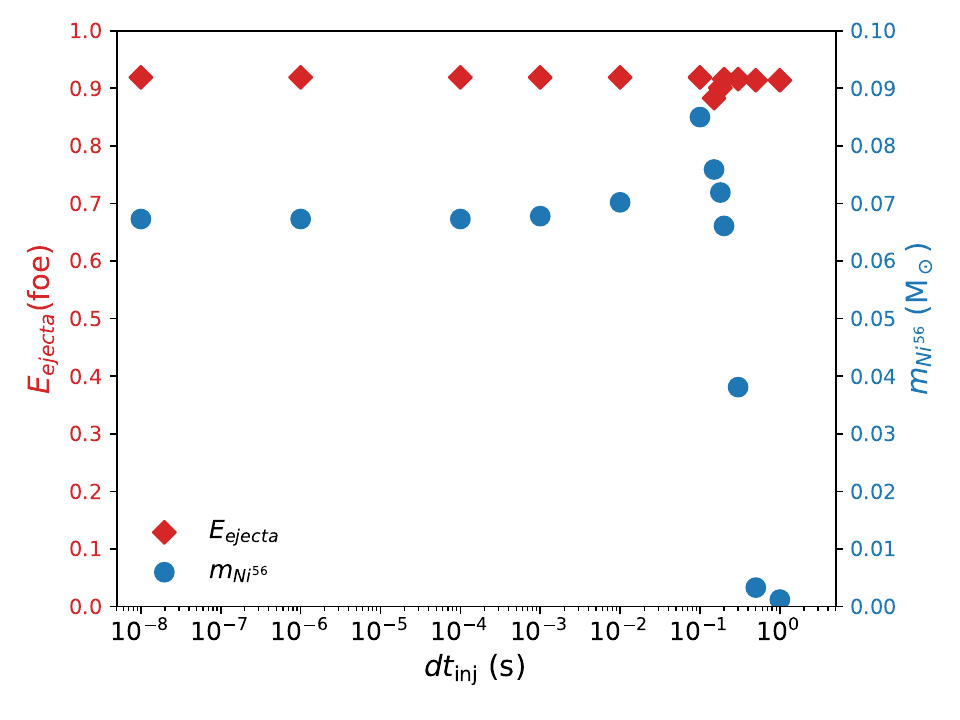}
    \caption{The Figure shows the values of $\Eejecta$ (red diamonds, whose scale is on the left-hand y-axes) and $\mnickel$ (blue points, whose scale is on the right-hand y-axis) as function of $dt_{\rm inj}$.}
    \label{fig_app: Calibration dt_inj}
\end{figure}

Figure \ref{fig_app: Calibration dt_inj} shows the values that we find for $\Eejecta$ and $\mnickel$ for different values of $dm_{\rm inj}$. We studied the variation of the outcomes of the explosion assuming $dt_{\rm inj}=10^{-9}s$ and $dE_{\rm inj}=2.0 \rm foe$, and by injecting the thermal energy at the arbitrary mass coordinate of $0.8\Ms$. The values we find for $\Eejecta$ and $\mnickel$ are also reported in Table \ref{tab_app: Calibration dt_inj}.\par
we find that the $\Eejecta$ is roughly independent on the timescale over which we inject the energy. On the other hand, the amount of $\nickel$ produced (i.e. the efficiency of the explosive Si-burning) is strongly dependent on $dt_{\rm inj}$. From Figure \ref{fig_app: Calibration dt_inj} we see that $\mnickel$ is constant for $dt_{\rm inj}\leq 10^{-3} s$, while for longer timescales the efficiency of the explosive nucleosynthesis increases, resulting in an overproduction of $\nickel$ with respect to SN1987A. The value of $\mnickel$ reaches its maximum for $dt_{\rm inj}=0.1s$, with $\mnickel\sim 0.085\Ms$, while on longer timescales it start decreasing.
This is because the shockwave reaches the Si shell, where the production of $\nickel$ takes place, at $\sim 0.2s $, thus if the thermal energy is not completely injected before this moment the shockwave will not have enough energy to trigger the complete explosive Si-burning. This is apparent from Figure \ref{fig_app: Calibration dt_inj} because for $dt_{\rm inj}\geq0.2s$ the amount of $\nickel$ produced is about an order of magnitude lower than what we obtain for lower $dt_{\rm inj}$.
Finally, the values of $dt_{\rm inj}$ that is closer to the observative properties of SN1987A is $dt_{\rm inj}=0.01s$, for which we obtain exactly $\mnickel=0.07\Ms$, which is the value one usually can find in the literature (e.g.  \citealt{Arnett1989, Shigeyama1990, Utrobin1993, Utrobin2006, Blinnikov2000}).

\section{Tables}

\begin{table*}
\caption{Results of the interpolation described in Sec. \ref{sec: PI regime}.  Column 1: Metallicity, Column 2: $\Mzam$ of the first progenitor undergoing PPISN, Column 3: Mass Pre-SN of the first progenitor undergoing PPISN. If the most massive progenitor of our grid is stable against PPISN we cannot obtain such information, thus we labeled it as Not Available (NA), Column 4: slope of Eq. \ref{eq: Mzams interpolation}, Column 5: intercept of Eq. \ref{eq: Mzams interpolation}, Column 6: slope of Eq. \ref{eq: Mpresn interpolation}, Column 7: intercept of Eq. \ref{eq: Mpresn interpolation}.}
\label{Tab: Interpolation results}
\centering
\begin{tabular}{c c c c c c c}
    \hline \hline
    $\met$ & $\Mzam^{\rm PPISN} (\Ms)$ & $M_{\rm preSN}^{\rm PPISN} (\Ms)$ & $k_1$ & $q_1 (\Ms)$ & $k_2$  & $q_2(\Ms)$ \\
    \hline
    \multicolumn{7}{c}{v=0~km/s}\\
    \hline
    0   & N.A.  & N.A.   & N.A. & N.A.  & N.A. & N.A. \\
    -1  & 87.43 & 41.91  & 0.28 & 8.3   & 0.27 & 18.3 \\
    -2  & 87.66 & 83.27  & 0.53 & -13.9 & 0.61 & 29.8 \\
    -3  & 87.01 & 87.01  & 0.48 & -9.2  & 1.0  & 0.0  \\
    \hline
    \multicolumn{7}{c}{v=300~km/s}\\
    \hline    
    0   & N.A. & N.A.    & N.A. & N.A.  & N.A. & N.A. \\
    -1  & 120.0 & 40.5   & 0.18 & 11.1  & 0.21 & 15.3 \\
    -2  & 65.19 & 39.89  & 0.39 & 7.9   & 0.48 & 8.6  \\
    -3  & 64.8  & 41.63  & 0.37 & -0.6  & 0.37 & 8.7  \\
    \hline
\end{tabular}

\end{table*}

\begin{table*}
    \caption{Final properties of stars with various initial metallicities and initial masses for the non-rotating stellar progenitors.  Column 1:  $\Mzam$, Column 2: Pre-SN helium core mass, Column 3: Pre-SN CO core mass, Column 4: Pre-SN mass, Column 5: Pre-SN binding energy, Column 6: Pre-SN total energy, Column 7: Final mass of the compact remnant, Column 8: Fraction of mass expelled during the explosion.}
    \centering
    \label{Tab: Non rotating simulations}
    \begin{tabular}{lccccccc}
    \hline\hline
    $\Mzam (\Ms)$ & $M_{\rm He}(\Ms)$& $M_{\rm CO}(\Ms)$ & $M_{\rm preSN}(\Ms)$ & $E_{\rm bind} (\rm foe)$ & $E_{\rm tot}(\rm foe)$  & $M_{\rm rem}(\Ms)$ & $dM/M$ \\
        \hline
        \multicolumn{8}{c}{[Fe/H]=-0~~~~v=0~km/s}\\
        \hline
         13   & 4.08  & 2.03  & 11.9    & 2.02    & -0.93     & 0.81  & 0.93\\
         15   & 4.95  & 2.78  & 13.2    & 2.47    & -1.07     & 0.81  & 0.94\\
         20   & 7.29  & 3.86  & 7.54    & 2.59    & -1.05     & 2.21  & 0.71\\
         25   & 8.54  & 6.21  & 8.54    & 3.58    & -1.35     & 3.69  & 0.57\\
         30   & 10.8  & 7.91  & 10.8    & 6.39    & -2.13     & 8.7   & 0.19 \\
         40   & 14.1  & 10.6  & 14.1    & 7.60    & -2.31     & 12.3  & 0.13\\
         60   & 16.9  & 13.0  & 16.9    & 10.6    & -3.04     & 16.2  & 0.04\\
         80   & 22.7  & 18.0  & 22.7    & 17.3    & -4.52     & 22.7  & 0.00\\
         120  & 27.9  & 22.0  & 27.9    & 24.4    & -6.07     & 27.9  & 0.00 \\
        \hline
        \multicolumn{8}{c}{[Fe/H]=-1~~~~v=0~km/s}\\
        \hline
         13  & 4.26  & 2.13  & 12.5     & 1.95    & -0.89         & 0.81  & 0.94 \\
         15  & 5.22  & 3.01  & 14.2     & 2.73    & -1.15         & 1.66  & 0.88 \\
         20  & 7.52  & 4.21  & 18.3     & 3.19    & -1.26         & 3.31  & 0.82 \\
         25  & 10.2  & 6.82  & 20.6     & 5.52    & -1.91         & 7.49  & 0.64 \\
         30  & 11.9  & 7.22  & 28.3     & 5.95    & -1.99         & 11.8  & 0.58 \\
         40  & 16.7  & 10.9  & 28.7     & 8.03    & -2.39         & 16.7  & 0.42 \\
         60  & 26.8  & 19.6  & 42.0     & 19.5    & -4.94         & 36.7  & 0.13 \\
         80  & 39.2  & 30.9  & 39.9     & 40.0    & -9.27         & 39.9  & 0.00 \\
         \hline
         \multicolumn{8}{c}{[Fe/H]=-2~~~~v=0~km/s}\\
         \hline
         13  & 4.34  & 2.14  & 13      & 2.25     & -1.01         & 0.81  & 0.94 \\
         15  & 5.21  & 2.72  & 14.8    & 2.25     & -0.95         & 0.82  & 0.94 \\
         20  & 7.49  & 4.23  & 19.7    & 3.26     & -1.28         & 3.44  & 0.83 \\
         25  & 9.87  & 5.93  & 24.7    & 4.05     & -1.47         & 5.77  & 0.77 \\
         30  & 11.7  & 6.98  & 29.9    & 5.46     & -1.84         & 12.7  & 0.58 \\
         40  & 16.7  & 11.2  & 39.7    & 8.78     & -2.57         & 25.8  & 0.35 \\
         60  & 27.2  & 20.0  & 59.4    & 20.7     & -5.16         & 59.4  & 0.00 \\
         80  & 38.3  & 28.9  & 78.6    & 36.1     & -8.18         & 78.6  & 0.00 \\
         \hline
         \multicolumn{8}{c}{[Fe/H]=-3~~~~v=0~km/s}\\
         \hline
         13  & 4.22  & 2.15  & 13     & 2.03      & -0.92         & 0.81 & 0.94  \\
         15  & 5.22  & 3.09  & 15     & 2.92      & -1.20         & 1.85 & 0.88 \\
         20  & 7.42  & 4.35  & 20     & 3.29      & -1.29         & 3.37 & 0.83 \\
         25  & 9.84  & 6.29  & 25     & 4.39      & -1.56         & 6.30 & 0.75 \\
         30  & 12.3  & 8.08  & 30     & 5.71      & -1.87         & 12.7 & 0.58 \\
         40  & 17.5  & 12.2  & 40     & 10.3      & -2.99         & 32.9 & 0.18 \\
         60  & 28.5  & 21.7  & 60     & 23.6      & -5.80         & 60   & 0.00 \\
         80  & 38.9  & 29.6  & 80     & 37.9      & -8.56         & 80   & 0.00 \\
    \hline
    \end{tabular}

\end{table*}

\begin{table*}
    \caption{Final properties of stars with various initial metallicities and initial masses for the non-rotating stellar progenitors.  Column 1:  $\Mzam$, Column 2: Pre-SN helium core mass, Column 3: Pre-SN CO core mass, Column 4: Pre-SN mass, Column 5: Pre-SN binding energy, Column 6: Pre-SN total energy, Column 7: Final mass of the compact remnant, Column 8: Fraction of mass expelled during the explosion.}
    \label{Tab: Rotating simulations}
    \centering
    \begin{tabular}{lccccccc}
    \hline\hline
    $\Mzam (\Ms)$ & $M_{\rm He}(\Ms)$& $M_{\rm CO}(\Ms)$ & $M_{\rm preSN}(\Ms)$ & $E_{\rm bind} (\rm foe)$ & $E_{\rm tot}(\rm foe)$  & $M_{\rm rem}(\Ms)$ & $dM/M$ \\
        \hline
        \multicolumn{8}{c}{[Fe/H]=-0~~~~v=300~km/s}\\
        \hline
         15  & 6.23   & 4.31   & 6.23   & 3.95   & -1.65   & 4.29 & 0.31 \\
        20  & 8.18   & 5.92   & 8.18   & 7.61   & -3.16   & 8.18 & 0.00 \\
        25  & 9.48   & 7.16   & 9.48   & 6.30   & -2.39   & 8.45 & 0.11 \\
        30  & 11.2   & 8.46   & 11.2   & 7.07   & -2.69   & 11.2 & 0.00 \\
        40  & 13.8   & 10.6   & 13.8   & 9.14   & -3.06   & 13.8 & 0.00 \\
        60  & 16.6   & 13.0   & 16.6   & 11.0   & -3.19   & 16.6 & 0.00 \\
        80  & 17.5   & 13.7   & 17.5   & 11.6   & -3.29   & 17.5 & 0.00 \\
        120 & 18.6   & 14.6   & 18.6   & 13.1   & -3.63   & 18.6 & 0.00 \\
        \hline
        \multicolumn{8}{c}{[Fe/H]=-1~~~~v=300~km/s}\\
        \hline
        15  & 6.50  & 4.89   & 11.6   & 5.85    & -2.77   & 11.6  & 0.00 \\
        20  & 8.12  & 6.16   & 17.1   & 7.21    & -3.21   & 17.1  & 0.00 \\
        25  & 12.2  & 9.73   & 18.5   & 8.52    & -3.07   & 12.6  & 0.32 \\
        30  & 16.0  & 12.3   & 16.0   & 11.9    & -3.85   & 16.0  & 0.00 \\
        40  & 20.7  & 16.2   & 20.7   & 16.3    & -4.7    & 20.7  & 0.00 \\
        60  & 27.5  & 21.4   & 27.5   & 23.7    & -6.21   & 27.5  & 0.00 \\
        80  & 32.1  & 25.7   & 32.1   & 29.8    & -7.25   & 32.1  & 0.00 \\
        120 & 40.5  & 33.0   & 40.5   & 41.8     & -9.33     & 40.5  & 0.00 \\
        \hline
        \multicolumn{8}{c}{[Fe/H]=-2~~~~v=300~km/s}\\
        \hline
        15  & 6.47  & 4.72  & 13.75   & 5.85     & -2.8    & 13.75 & 0.00 \\
        20  & 9.85  & 6.77  & 16.8    & 8.83     & -3.45   & 15.3  & 0.09 \\
        25  & 13.2  & 10.4  & 13.2    & 9.37     & -3.28   & 13.2  & 0.00 \\
        30  & 15.1  & 11.7  & 15.6    & 10.5     & -3.35   & 15.6  & 0.00 \\
        40  & 17.6  & 18.5  & 22.9    & 22.9     & -5.24   & 22.9  & 0.00 \\
        60  & 37.4  & 31.0  & 37.4    & 42.7     & -11.0   & 37.4  & 0.00 \\
        \hline
        \multicolumn{8}{c}{[Fe/H]=-3~~~~v=300~km/s}\\
        \hline
        15  & 6.09  & 4.16   & 13.8  & 4.23      & -1.81    & 4.38 & 0.68 \\
        20  & 7.89  & 5.29   & 20    & 7.28      & -3.1     & 20.0 & 0.00 \\
        25  & 13.0  & 10.5   & 13.3  & 9.41      & -3.22    & 13.3 & 0.00 \\
        30  & 16.9  & 13.1   & 17.1  & 11.2      & -3.68    & 17.1 & 0.00 \\
        40  & 24.5  & 19.1   & 24.5  & 20.1      & -5.72    & 24.5 & 0.00 \\
        60  & 38.1  & 29.4   & 38.1  & 37.1      & -9.24    & 38.1 & 0.00 \\
    \hline
    \end{tabular}

\end{table*}

\section{Calibration Tables}

\begin{table}
\centering
\caption{Calibration on $E_{\rm inj}$. Column 1: $E_{\rm inj}$, Column 2: $\Eejecta$ of the explosion, Column 3: $\mnickel$ produced in the explosive nucleosynthesis, Column 4: $M_{rem}$ we find for the explosion.}
\label{tab_app: Calibration E_inj}
    \begin{tabular}{lccc}
        \hline \hline
        $E_{\rm inj} (\rm foe)$ & $\Eejecta (\rm foe)$ & $\mnickel (\Ms)$ & $M_{rem} (\Ms)$\\
        \hline
        1.9  & 0.82   & 0.059  & 0.835    \\
        2.0  & 0.92   & 0.067  & 0.829    \\
        2.1  & 0.99   & 0.076  & 0.822    \\
        \hline
    \end{tabular}
\end{table}

\begin{table}
    \centering
    \caption{Calibration on $dm_{\rm inj}$. Column 1: $dm_{\rm inj}$, Column 2: $\Eejecta$ of the explosion, Column 3: $\mnickel$ produced in the explosive nucleosynthesis, Column 4: $M_{rem}$ we find for the explosion.}
    \label{tab_app: Calibration dm_inj}
    \begin{tabular}{lccc}
        \hline \hline
        $E_{\rm inj} (\rm foe)$ & $\Eejecta (\rm foe)$ & $\mnickel (\Ms)$ & $M_{rem} (\Ms)$\\
        \hline
        0.400  & 0.948  & 0.073    & 1.148    \\
        0.200  & 1.070  & 0.116    & 0.923   \\
        0.100  & 0.919  & 0.067    & 0.829   \\
        0.050  & 0.995  & 0.111    & 0.858   \\
        0.010  & 0.999  & 0.109    & 0.879   \\
        0.005  & 0.999  & 0.108    & 0.897   \\
        0.001  & 0.997  & 0.104    & 0.905   \\
        \hline
    \end{tabular}
\end{table}

\begin{table}
\centering
\caption{Calibration on $dt_{\rm inj}$. Column 1: $dt_{\rm inj}$, Column 2: $\Eejecta$ of the explosion, Column 3: $\mnickel$ produced in the explosive nucleosynthesis, Column 4: $M_{rem}$ we find for the explosion.}
\label{tab_app: Calibration dt_inj}
\begin{tabular}{lccc}
\hline \hline
$dt_{\rm inj} (\Ms)$ & $\Eejecta (\rm foe)$ & $\mnickel (\Ms)$ & $M_{rem} (\Ms)$ \\
\hline
$1 \times 10^0$      & 0.914 & 0.001 & 0.801   \\
$5 \times 10^{-1}$   & 0.914 & 0.003 & 0.801   \\
$3 \times 10^{-1}$   & 0.916 & 0.004 & 0.815   \\
$2 \times 10^{-1}$   & 0.916 & 0.066 & 0.801   \\
$1.8 \times 10^{-1}$ & 0.901 & 0.072 & 0.808   \\
$1.5 \times 10^{-1}$ & 0.883 & 0.076 & 0.822   \\
$1 \times 10^{-1}$   & 0.919 & 0.085 & 0.822   \\
$1 \times 10^{-2}$   & 0.919 & 0.070 & 0.822   \\
$1 \times 10^{-3}$   & 0.919 & 0.068 & 0.829   \\
$1 \times 10^{-4}$   & 0.919 & 0.067 & 0.829   \\
$1 \times 10^{-6}$   & 0.919 & 0.067 & 0.829   \\
$1 \times 10^{-8}$   & 0.919 & 0.067 & 0.822   \\
\hline
\end{tabular}

\end{table}

\end{document}